\documentclass[
  journal=largetwo,
  manuscript=article-type,
  year=2020,
  volume=37,
]{cup-journal}
\usepackage{aas_macros}
\usepackage{amsmath}
\usepackage[nopatch]{microtype}
\usepackage{booktabs}
\usepackage{graphicx}	
\usepackage{amsmath}	
\usepackage{color,soul}
\usepackage{longtable}
\usepackage{pdflscape}
\DeclareUnicodeCharacter{0301}{\hspace{-1ex}\'{ }}

\usepackage[breaklinks=true]{hyperref}
\hypersetup{colorlinks,linkcolor=blue,citecolor=blue,urlcolor=blue}
\usepackage{graphicx}
\usepackage{txfonts}
\usepackage{lscape}
\usepackage{longtable}
\usepackage{multirow}
\usepackage{multicol}
\usepackage{amssymb}
\usepackage{pifont}
\usepackage[export]{adjustbox}

\title{Searching for stellar population in the molecular cloud GRSMC~045.49+00.04}

\author{N. Azatyan}
\affiliation{Byurakan Astrophysical Observatory, 0213 Aragatsotn Prov., Armenia}
\email[N. Azatyan]{nayazatyan@bao.sci.am}

\author{E. Nikoghosayn}
\affiliation{Byurakan Astrophysical Observatory, 0213 Aragatsotn Prov., Armenia}

\author{A. Samsonyan}
\affiliation{Byurakan Astrophysical Observatory, 0213 Aragatsotn Prov., Armenia}

\author{D. Elia}
\affiliation{NAF – Istituto di Astrofisica e Planetologia Spaziali, Via Fosso del Cavaliere 100, I-00133 Roma, Italy}

\author{L. Kaper}
\affiliation{Anton Pannekoek Institute, University of Amsterdam, Science Park 904, 1098 XH Amsterdam, The Netherlands}

\author{A. Yeghikyan}
\affiliation{Byurakan Astrophysical Observatory, 0213 Aragatsotn Prov., Armenia}

\author{D. Andreasyan}
\affiliation{Byurakan Astrophysical Observatory, 0213 Aragatsotn Prov., Armenia}

\author{D. Baghdasaryan}
\affiliation{Byurakan Astrophysical Observatory, 0213 Aragatsotn Prov., Armenia}


\begin{document}

\begin{abstract}
Understanding the characteristics of young stellar populations is essential for deriving insights into star formation processes within parent molecular clouds and the influence of massive stars on these processes. This study primarily aims to investigate the young stellar objects (YSOs) within the molecular cloud G\,045.49+00.04, including three ultra-compact HII (UC\,HII) regions: G\,45.48+0.13 (IRAS\,19117+1107), G\,45.45+0.06 (IRAS\,19120+1103), and G\,45.47+0.05. We used near-, mid-, and far-infrared photometric data along with radiation transfer models and the modified blackbody fitting to identify and study the YSOs and the interstellar medium (ISM). In total, we identified 1482 YSOs in a 12 arcmin radius covering GRSMC\,045.49+00.04, with a mass range from 1.5 M$_{\odot}$ to 22 M$_{\odot}$. Of these, 315 objects form relatively dense clusters in the UC\,HII regions, close to the IRAS\,19120+1103 and 19117+1107 sources. In each UC\,HII region, several high-mass stars have been identified, which in all likelihood are responsible for the ionization. The YSOs with 21.8\,M$_{\odot}$ and 13.7\,$\pm$ 0.4\,M$_{\odot}$ are associated with IRAS\,19120+1103 and 19117+1107, respectively. The non-cluster YSOs (1168) are uniformly distributed on the field. The distribution of YSOs from both samples on the colour-magnitude diagram and by the evolutionary ages is different. About 75\% of objects in the IRAS clusters are concentrated around the Zero Age Main Sequence and have a well-defined peak at an age of Log(Age[years])\,$\approx$\,6.75, with a narrow spread. The non-cluster objects have two concentrations located to the right and left of the 0.1 Myr isochrone and two well-defined peaks at Log(Age)\,$\approx$\,6.25 and 5.25. The fraction of the near-infrared excess stars, as well as the mass function confirm that the evolutionary age of the cluster is about 1\,Myr. The K luminosity functions' $\alpha$ slopes for the IRAS clusters and non-cluster objects are 0.32$\pm$0.04 and 0.72$\pm$0.13, respectively. The steeper $\alpha$ slope is suggesting that the non-cluster objects are less evolved, which is well consistent with the evolutionary age. Similar results - including evolutionary age, narrow age spread, and the less evolved nature of non-cluster objects - were also observed for the YSOs in the neighboring   G\,45.14+00.14. The both regions (G\,045.49+00.04 and G\,45.14+00.14) are located and distinguished by their brightness and density at the edge of the bubble around the highly variable X-ray binary GRS\,1915+105, which includes a black hole and a K-giant companion. Based on the above, we can assume that the process of star formation in the young IRAS clusters was triggered by the GRS\,1915+105-initiated shock front inside the ISM massive condensation, through the process of "collecting and collapse". Most non-cluster objects probably belong to a later generation. Their formation could be triggered by the recurrent activity of GRS\,1915+105 and/or through the edge collapse scenario and mass accumulation through the gas flows along the ISM filaments.  

\end{abstract}

\section{Introduction}
\label{sec:introduction}
Currently, it is widely accepted that massive star formation takes place within dense ($n>$\,10$^3$\,cm$^{-3}$) and cold ($T\sim$\,10\,--\,50\,K) clumps found in giant molecular clouds \citep{Lada2003,Ascenso2018}. Young stars form on a wide range of scales, producing clusters with various degrees of gravitational self-binding, mass segregation, density, etc. \citep{Elmegreen2000}. In general, young clusters have a hierarchical structure both in space and in time; that is, there may be groups of young stars in different stages of evolution. 

Massive stars ($M\geqslant$\,8\,$M_{\odot}$) play a key role in the star formation process. They provide external pressure in the form of expanding HII regions, stellar winds, supernovae explosions, and powerful outflows in the surrounding gas. Therefore, they can sustain sequential and self-propagating star formation processes \citep{Zinnecker2007}. Achieving a complete theoretical understanding of the formation and evolution of massive stars is a major component in developing a general  star formation theory, which seeks to explain the birth of stars of all masses in various star-forming environments.

The Lyman continuum emission from massive stars ionizes their immediate environment, forming HII regions. These regions are theorized to evolve through several stages, starting from hyper-compact to ultra-compact (UC)\,HII, then progressing to compact, and eventually to diffuse phases. Throughout these stages, they might either dissociate or clear away the surrounding medium \citep{Churchwell2002, Keto2007}. UC\,HII regions are particularly noteworthy as they mark the initial phases of massive star formation and provide critical "laboratories" for investigating the effects of these stars on their surroundings and the triggering of subsequent star formation cycles. Studies have demonstrated that UC\,HIIs are not exclusively associated with individual young massive stars; they often occur around clusters of such stars \citep[e.g.][]{Feldt1998, Andreasyan2020, Azatyan2020, Azatyan22}.
Therefore, the study of such formations is of great importance for understanding the evolutionary development of the massive stars themselves, their impact on the environment, and the theory of star formation in general.

\begin{figure}
	\includegraphics[width=0.95\columnwidth]{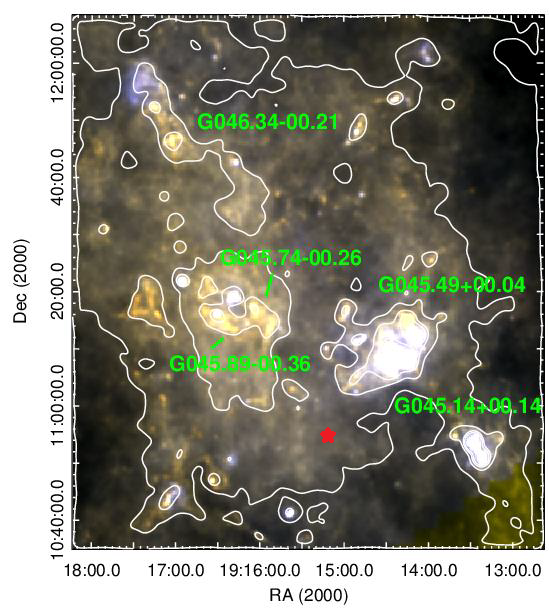}
    \caption{Colour-composite \textit{Herschel} image of GRSMC\,45.46+0.05 molecular cloud: 160\,$\mu$m (blue), 350\,$\mu$m (green), and 500\,$\mu$m (red) bands. The location of GRS\,1915+105 is indicated by a red star.}
    \label{fig:Common}
\end{figure}

The aim of our research is to search for and determine the
main properties of a young stellar population within the molecular cloud G\,045.49+00.04, including mass, evolutionary age and age spread, spatial distribution, luminosity and mass functions. This molecular cloud is a part of the massive and UC\,HII-rich star-forming region known as the Galactic Ring Survey Molecular Cloud (GRSMC) 45.46+0.05, located close to the Galactic plane \citep{Rathborne2009}. The position of G\,045.49+00.04 within GRSMC\,45.46+0.05 is shown in Fig.~\ref{fig:Common}.

\citet{Pandian2009} calculated, using methanol masers and a flat rotation curve, a range of distances for the clumps within G\,45.49+00.04, from 6.9 to 7.4\,kpc. This aligns well with the estimate from \citet{Roman2009} who used HI self-absorption and continuum absorption to determine a distance of $\sim$\,7.0\,$\pm$\,0.5\,kpc, as well as with the measurements of trigonometric parallaxes and proper motions of molecular masers \citep[7.75\,$\pm$\,0.45\,kpc,][]{Wu2019}. The latter distance value will be used for this study. To the south of G\,45.49+00.04, at almost the same distance within the error bar (8.6$_{-1.6}^{+2.0}$ kpc), lies the high-mass X-ray binary GRS\,1915+105, which contains a black hole and a K-giant companion. The black hole mass calculated for this distance is 12.4$_{-1.8}^{+2.0}$\,M$_{\odot}$ \citep{Reid2014,Motta2021}.

The G\,45.49+00.04 cloud is associated with different manifestations of star-forming activity. In particular, a number of studies have identified maser emissions in this region: H$_2$O \citep{Kim2019}, OH \citep{Engels2007}, SiO  \citep{Jewell1991}, and CH$_3$OH at a frequency of 6.7\,GHz \citep{Pandian2007,Pestalozzi2005}.

\begin{figure}
	\includegraphics[width=0.93\columnwidth]{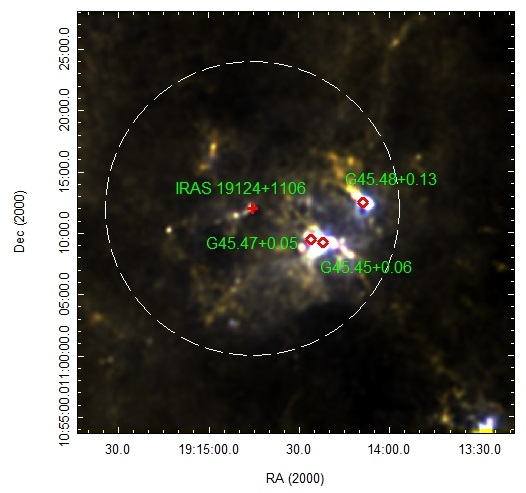}
    \caption{Colour-composite \textit{Herschel} image of GRSMC\,045.49+00.04 molecular cloud: 160\,$\mu$m (blue), 350\,$\mu$m (green), and 500\,$\mu$m (red) bands. The positions of UC\,HII regions are marked by red diamonds and IRAS source by red cross. The dashed white circle outlines an area with a radius of 12\'\, covering the designated study region.}
    \label{fig:Herschel}
\end{figure}

The G\,45.49+00.04 includes three UC\,HII regions (G\,45.48+0.13, G\,45.45+0.06 and G\,45.47+0.05) associated with the IRAS\,19117+1107 and IRAS\,19120+1103 sources and dense molecular cores selected from the Galactic Ring Survey in the mid-infrared \citep{Kraemer2003} (see Fig.~\ref{fig:Herschel}). According to \citet{Wood1989}, G\,45.48+0.13 (IRAS\,19117+1107) is an irregular or multiply-peaked UC\,HII region with size of 20"\,x\,4". The map of the radio continuum emission shows that this region has a bipolar structure with an angular size of $\sim$\,30" \citep{Garay1993}. IRAS\,19120+1103 is associated with other two UC\,HIIs. One of them (closer to the IRAS source), G\,45.45+0.06, is cometary-shaped with a diameter of approximately 7.4".  The other, G\,45.47+0.05, is an irregular or multiply-peaked region with a diameter of $\sim$\,3" \citep{Wood1989}. The map of the radio continuum emission shows that G\,45.45+0.06 region has core-halo structure with an angular size of $\sim$\,28" \citep{Garay1993}. \citet{Fuente202} have shown that G\,45.45+0.06 is directly connected with the surrounding extended emission. \citet{Cyganowski2008} revealed several extended 4.5\,$\mu$m emission objects in the vicinity of IRAS\,19120+1103 and 19117+1107, which may be traces of shocked molecular gas in a protostellar outflow.

The existence of young stellar objects (YSOs) in the vicinity of IRAS\,19120+1103 has been considered in several papers using near- and mid-infrared photometric data  \citep{Feldt1998,Blum2008,Paron2009}. Generally, the authors have concluded that several young massive O-type stars are responsible for the ionization of G\,45.45+0.06 region. In addition, several probable YSOs have been found, predominantly located in the molecular shell surrounding the core of the cloud. The photometric analysis of point-like sources by \citet{Bhadari2022} has shown clustering of YSO candidates in the vicinity of the UC\,HII regions.

In this paper, we present the results of a near-, mid-, and far-infrared (NIR, MIR, and FIR, respectively) study of the stellar population in the G\,045.49+00.04 cloud including all three UC\,HII regions. Our aim is to gain a better understanding of the properties of the embedded young stellar population and nature of their formation processes. The paper is organized as follows: Section \ref{sec:data} describes the data used; Section \ref{sec:results} presents the results obtained and their discussion: identification of YSOs (Sec. \ref{subsec:selection}); distributions of YSOs (Sec. \ref{sec:dis}), colour-magnitude diagram and evolutionary age spread (Sec. \ref{sec:cmd}); K luminosity function (Sec. \ref{sec:klf}), mass function (Sec. \ref{sec:IMF}), and a comparative analysis of young stellar populations in the G\,045.49+00.04 and neighboring G\,045.14+00.14 regions, along with an examination of potential triggers for star formation (Sec. \ref{sec:Trigger}). Finally, the conclusions drawn from the obtained results are presented in Section \ref{sec:con}.

\section{Used data}
 \label{sec:data}

As noted in Introduction, one of the objectives of this study is to conduct a comparative analysis of the properties of the young stellar population in the G\,045.49+00.04 region with the findings from the neighboring G\,045.14+00.14 region. Therefore, to identify and determine the main parameters of YSOs, we used the same archival observational data, as in the previous study of the G\,045.14+00.14 region \citep[][hereafter referred to as Paper\,I]{Azatyan22}. Specifically, for the study of YSOs, we employed NIR data from the United Kingdom Infrared Telescope Infrared Deep Sky Survey Galactic Plane Survey \citep[UKIDSS GPS;][]{lucas08}, as well as the MIR data from the Galactic Legacy Infrared Midplane Survey Extraordinaire \citep[GLIMPSE;][]{churchwell09}, the Multiband Infrared Photometer for \textit{Spitzer} (MIPSGAL) \citep{carey09}, and  the Wide-field Infrared Survey Explorer \citep[WISE;][]{wright10} databases. Additionally, we used FIR data to determine the parameters of both YSOs and the interstellar medium (ISM). The FIR data were sourced from the Photodetector Array Camera and Spectrometer \citep[PACS;][]{poglitsch10}, the Spectral and Photometric Imaging Receiver \citep[SPIRE;][]{griffin10}, and the Hi-GAL infrared Galactic Plane Survey \citep[Hi-GAL;][]{molinari16}.

\begin{figure}
	\includegraphics[width=0.93\columnwidth]{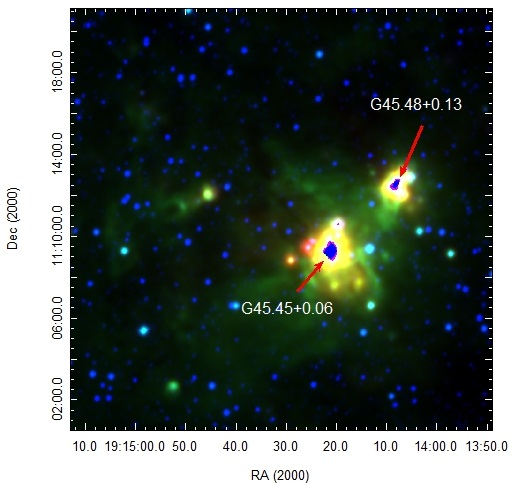}
    \caption{Colour-composite WISE image of GRSMC\,45.46+0.05 molecular cloud: W2 (blue), W3 (green), and W4 (red) bands. The positions of saturated regions are indicated by red arrows.}
    \label{fig:MIR}
\end{figure}


\section{Results and Discussion}
\label{sec:results}

\subsection{Selection of young stellar population}
\label{subsec:selection}

\begin{figure}[t]
	\includegraphics[width=\columnwidth]{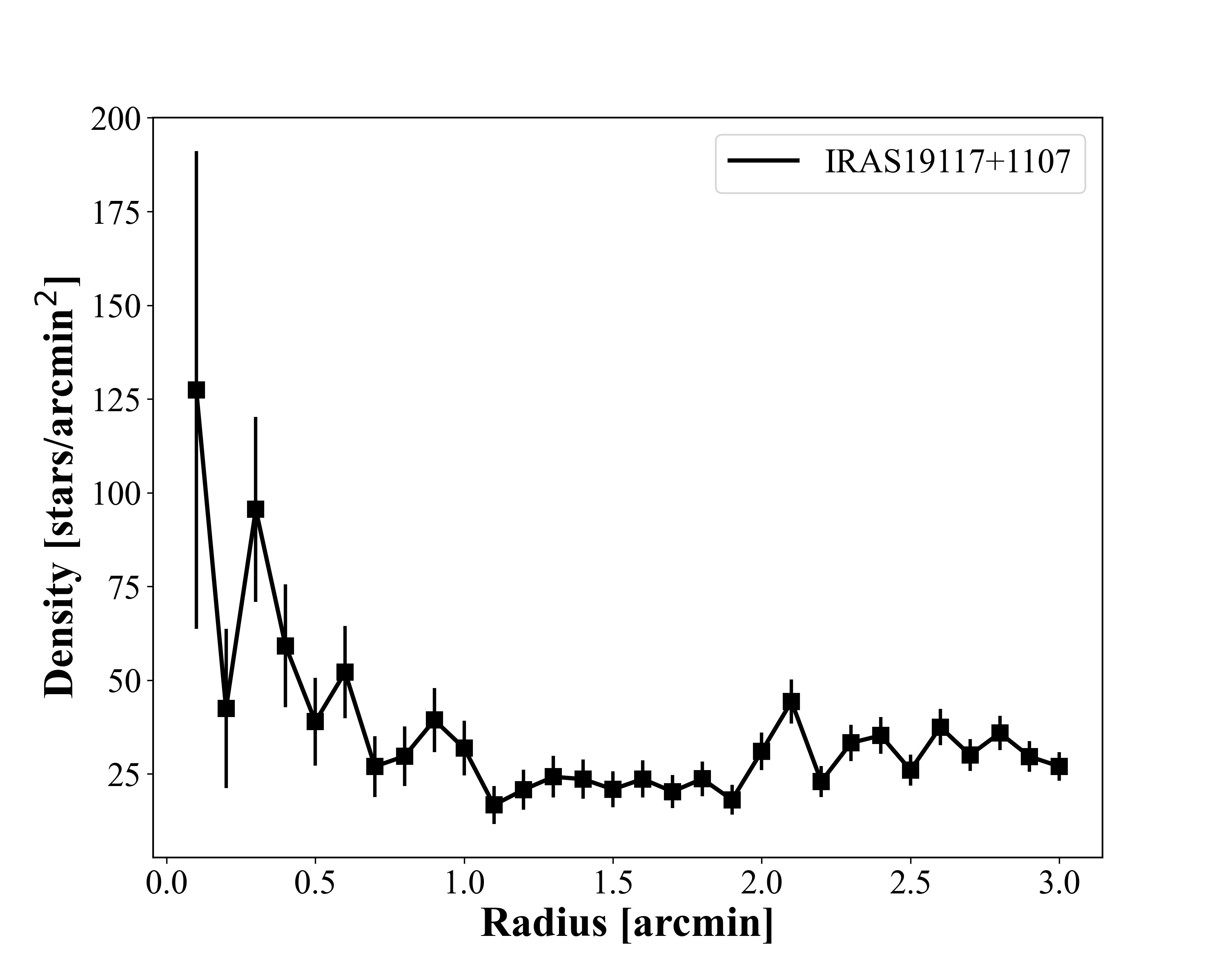}
	\includegraphics[width=\columnwidth]{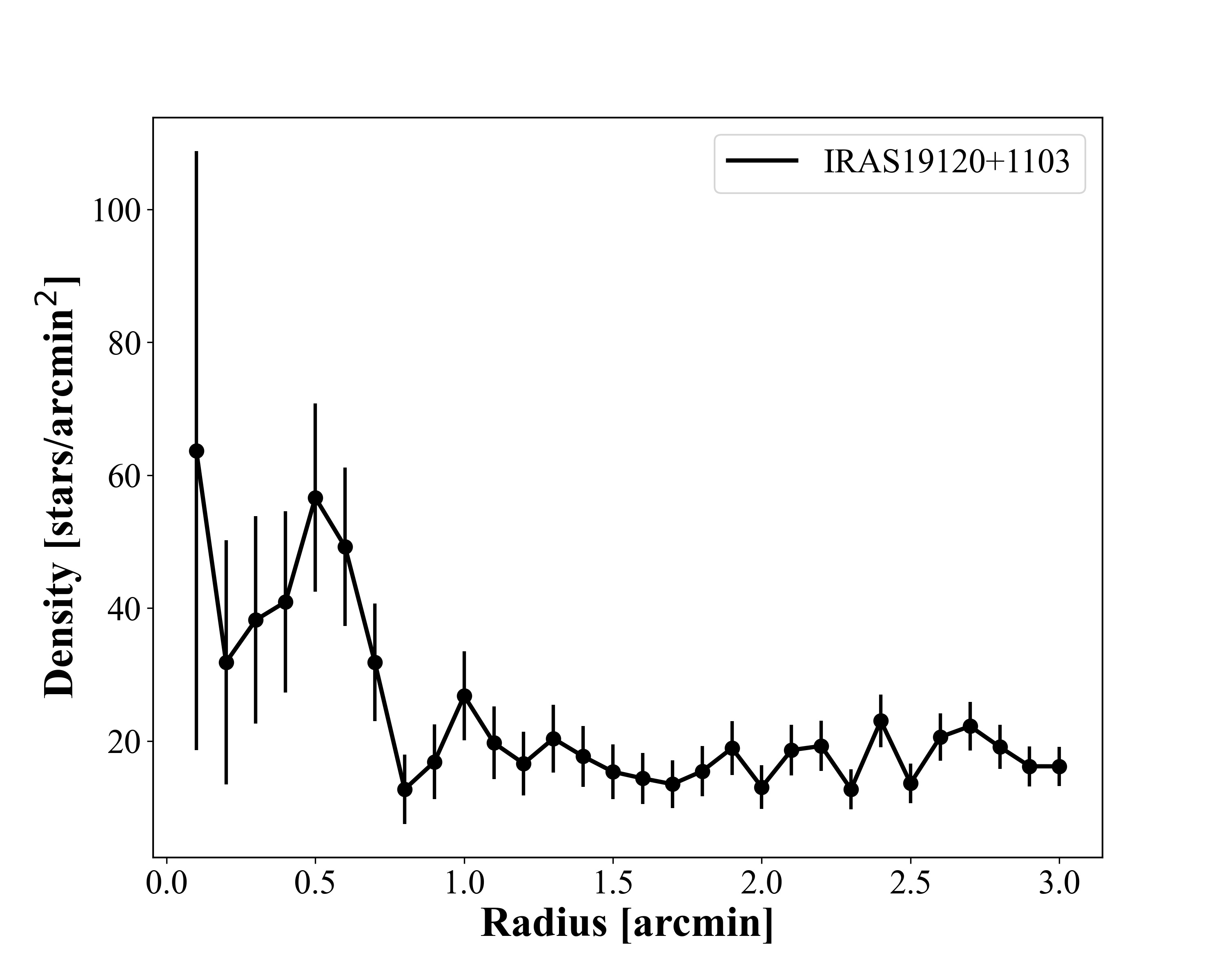}
    \caption{Radial distribution of the stellar surface density relative to the IRAS\,19117+1107 (top panel) and IRAS\,19120+1103 (bottom panel), respectively. Vertical lines \textbf{represent} standard errors. \textbf{The distributions are based on point sources from the GPS UKIDSS-DR6 catalogue.}}
    \label{fig:radial}
\end{figure}

Potential members of the star formation region were identified and examined within an area encompassing the entire G045.49+00.04 molecular cloud. This area is defined by a circle with a radius of 12\'\,, centered at the coordinates $\alpha$\,=\,19:14:45.3 and $\delta$\,=\,+11:11:58.2, as depicted in Fig.~\ref{fig:Herschel}. This area significantly exceeds the surface area of G\,045.49+00.04 clump, which was determined to be $\sim$\,0.02\,deg$^2$ in \citet{Rathborne2009}. The area includes three UC\,HII regions (G\,45.48+0.13, G\,45.45+0.06, and G\,45.47+0.05; see Sec.~\ref{sec:introduction}), which are marked. We also marked the position of IRAS\,19124+1106, postulated in the previous study to be associated with the termination in the ISM of the sub-arcsecond radio jet from black hole transient GRS\,1915+105 \citep{Rodriguez1998}. However, it should be noted that, as discussed by \citet{Zdziarski2005}, the association of IRAS\,19124+1106 with GRS\,1915+105 is highly uncertain. In total, in the selected area 106,000 point sources were identified from the GPS UKIDSS-DR6 catalogue. Figure \ref{fig:radial} presents the radial density distributions of point sources relative to the IRAS\,19117+1107 and 19120+1103 sources. The stellar density was calculated in rings of 0.1\'\ width, by dividing the number of stars by the surface area. The standard error for the number of stars in each ring served as the measure of uncertainty. It is evident that within radii of $\sim$\,1.1\'\ for IRAS\,19117+1107 and $\sim$\,1.5\'\ for IRAS\,19120+1103, the stellar density significantly exceeds the average field density, indicating the presence of stellar clusters within these radii.

To select potential cluster members from the 106,000 identified point sources, we proceeded from the assumption that overwhelming majority of members of the considered active star-forming regions are YSOs. One of the most common and significant observational indicators of YSOs is the infrared (IR) excess, attributed to the presence of circumstellar envelopes and disks \citep{Lada2003}. Additionally, the level of IR excess in the NIR and/or MIR ranges can be used to determine the evolutionary stage of a YSO (ranging from Class\,I to Class\,III). Consequently, the identification of potential stellar members within the star formation region relies on their IR photometric data (NIR, MIR, and FIR). The detailed selection procedure, outlined in Paper\,I, involves several key steps: (i) IR colour-colour (c-c) diagrams, (ii) extraction of field contamination, and (iii) Spectral Energy Distribution (SED) fitting through radiative transfer models.\\

\textit{Colour-colour diagrams}. The initial step involves using NIR and MIR colour indices, or c-c diagrams, to identify stellar objects exhibiting IR excess. Numerous studies have provided theoretical justifications that explain the placement of YSOs within the c-c diagrams, correlating specific positions with different evolutionary stages. For our analysis, as in Paper\,I, we employed six distinct c-c diagrams:

\begin{itemize}
    \item (J–H) vs. (H–K) c–c NIR diagram \citep{meyer97, hernandez05, Lada1992};
    \item K–[3.6] vs. [3.6]–[4.5] c–c diagram, which combines NIR and MIR photometric data \citep{allen07};
    \item \textit{Spitzer} MIR [3.6]-[4.5] vs. [5.8]-[8.0] and [3.6]-[5.8] vs. [8.0]-[24] c-c diagrams \citep{allen07};
    \item WISE MIR [3.4]-[4.6] vs. [4.6]-[12] and [3.4]-[4.6] vs. [4.6]-[22] diagrams \citep{Koenig2012}.
\end{itemize}

\textit{Extraction of field contamination}.  Aside from YSOs, other objects can also display an IR excess and may be mistakenly identified as YSOs. These objects include: (i) star-forming galaxies and narrow-line active galactic nuclei (AGNs), which exhibit increasing excesses at 5.8 and 8.0\,$\mu$m due to hydrocarbon emissions; (ii) broad-line AGNs, whose IRAC colours closely resemble those of YSOs; and (iii) Asymptotic Giant Branch (AGB) stars, which also present a significant IR excess. To maintain the purity of our YSO sample, we excluded all point sources with colour indices typical of the aforementioned categories, as identified by \citet{gutermuth08,qiu08,stern05} for galaxies and by \citet{robitaille08} for AGB stars.\\

\begin{figure*}[h!t]
	\includegraphics[width=0.3\linewidth]{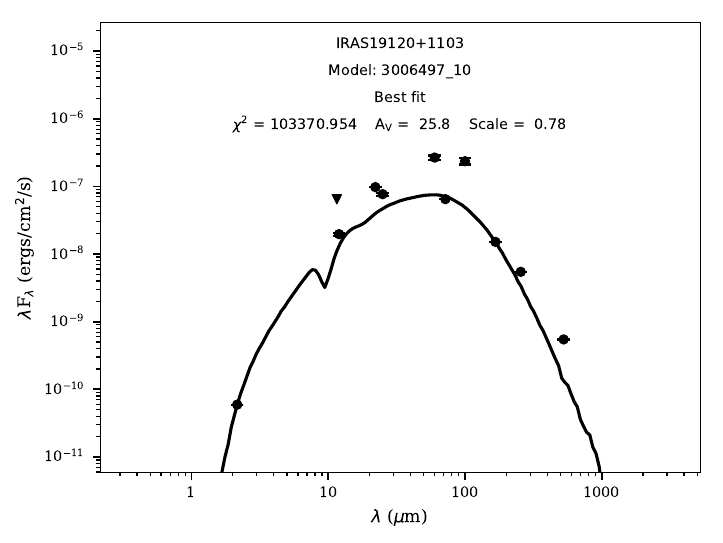}
	\includegraphics[width=0.3\linewidth]{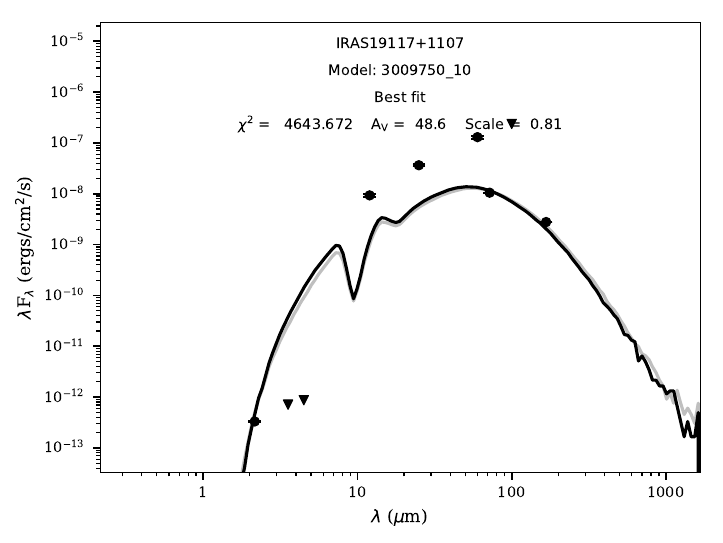}
	\includegraphics[width=0.3\linewidth]{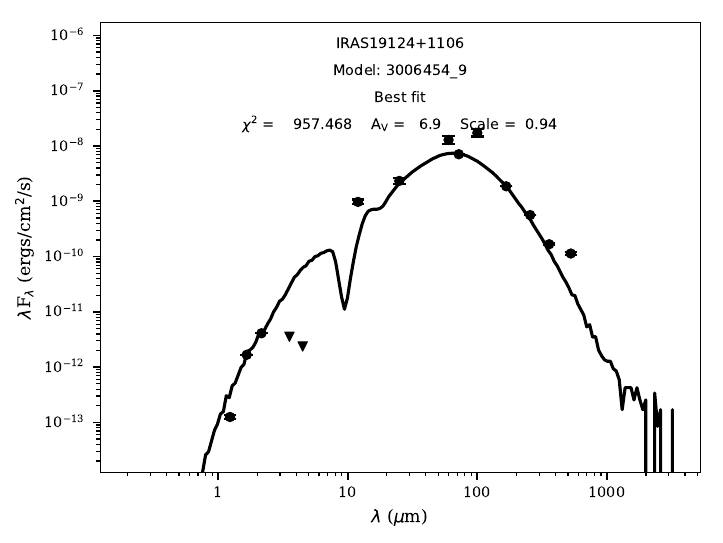}
    \includegraphics[width=0.3\linewidth]{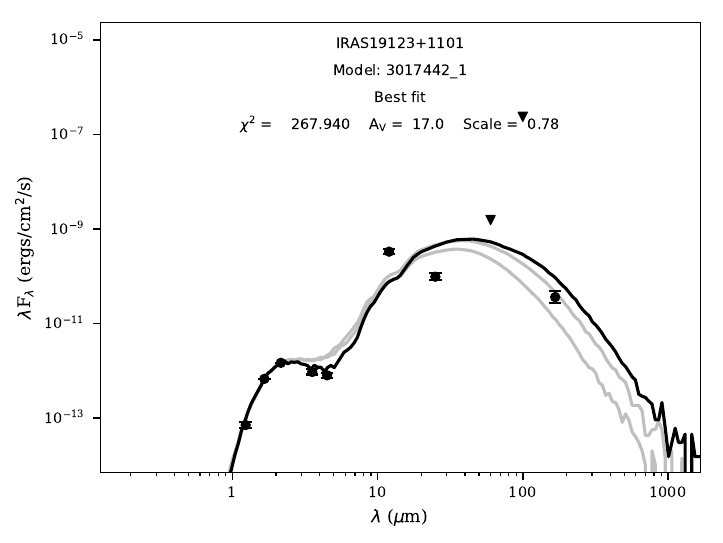}
	\includegraphics[width=0.3\linewidth]{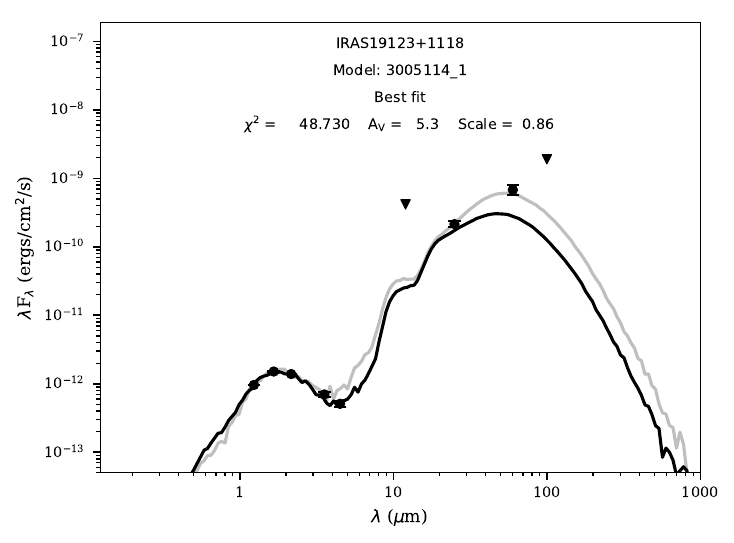}
	\includegraphics[width=0.3\linewidth]{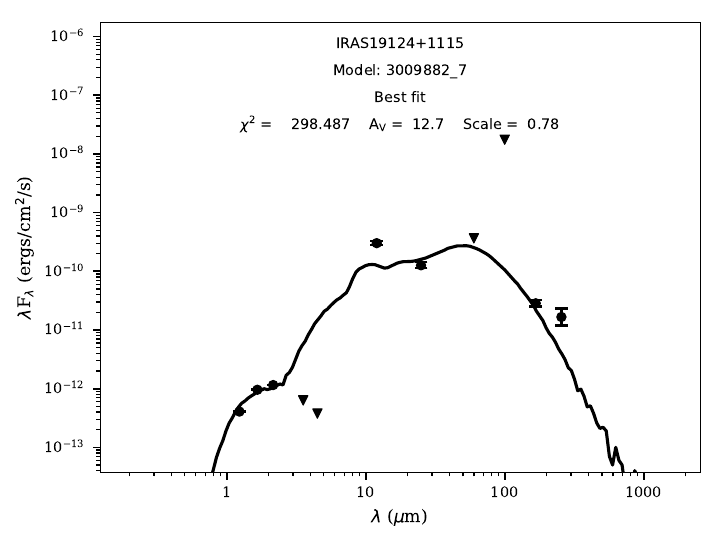}
	\includegraphics[width=0.3\linewidth]{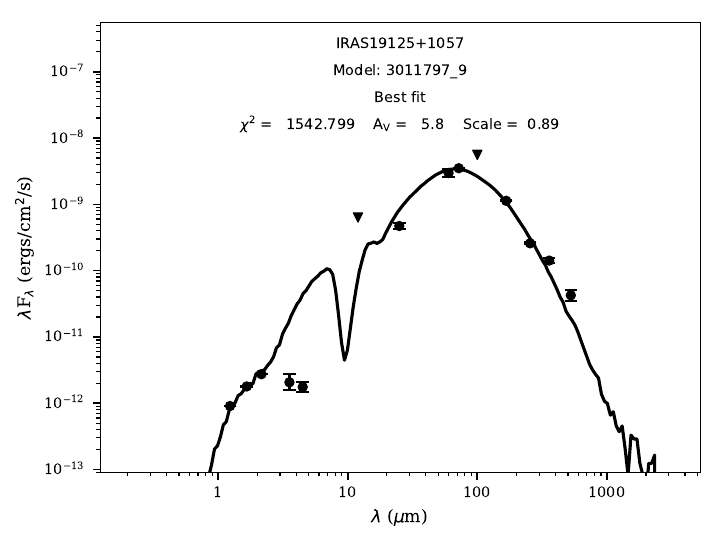}
	\includegraphics[width=0.3\linewidth]{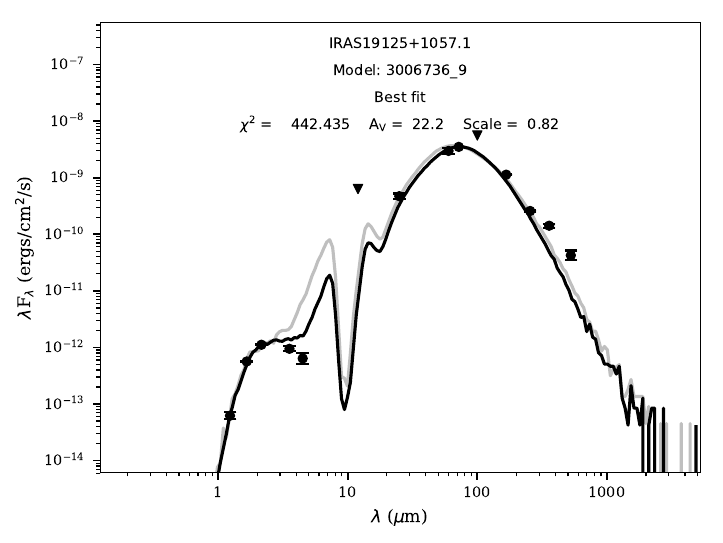}
	\includegraphics[width=0.3\linewidth]{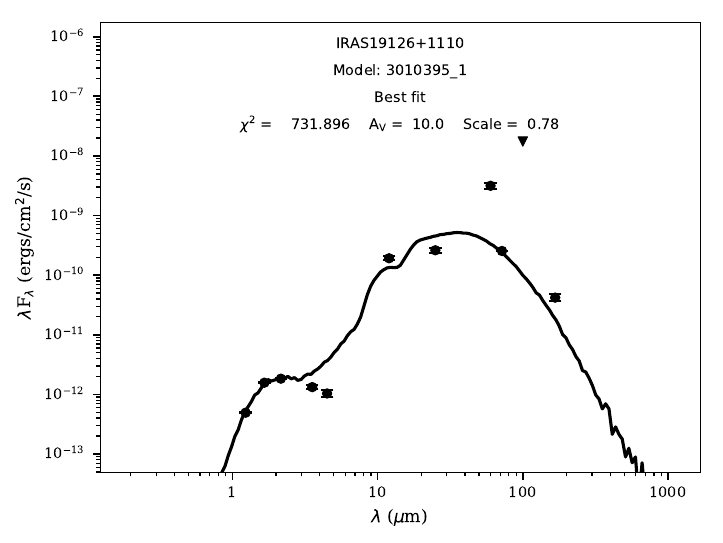}
	\includegraphics[width=0.3\linewidth]{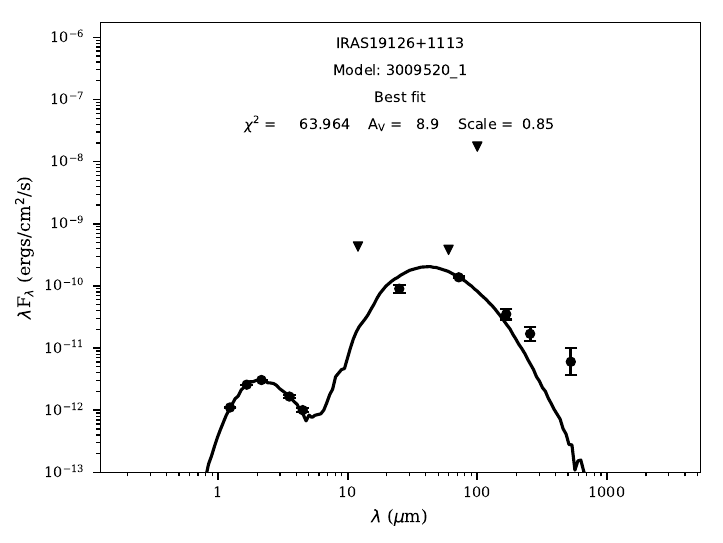}	
    \caption{Infrared SEDs of IRAS sources, which have been fitted to the models of \citet{robitaille07}. The filled circles represent the input fluxes, while triangles represent upper limits from the respective FIR clumps. The black line shows the best fitting model and gray lines show the subsequent good fits.}
    \label{fig:SED}
\end{figure*}

To minimize the risk of incorrect identification, we selected YSOs based on the criterion that they must be classified as objects with IR excess in at least two different c-c diagrams. However, the presence of two saturated areas in the MIR band near the IRAS objects ($\alpha$\,=\,19:14:07.13, $\delta$\,=\,+11:12:40.5 with a radius of $\sim$\,60" for IRAS\,19117+1107, and $\alpha$\,=\,19:14:21.37, $\delta$\,=\,+11:09:56.0 with a radius of $\sim$\,90" for IRAS\,19120+1103) may result in the loss of potential members (see Figure \ref{fig:MIR}). Since objects in the UC\,HII regions are of particular interest, we classified them as YSOs based solely on the NIR c-c diagram. Ultimately, we selected a total of 2,031 YSOs within a 12\'\ radius, comprising 1,874 Class\,II and 157 Class\,I objects. Of these, 217 were selected based exclusively on NIR data.\\

\textit{SED analysis.} To verify the selection of YSOs and to determine their parameters, including stellar mass and evolutionary age, we fitted their SEDs with the radiative transfer models developed by \citet{robitaille07}. This methodology is thoroughly explained in Paper I. For interstellar extinction, we selected a range of 5–100 mag, which is broader than the values from COBE/DIRBE and IRAS/ISSA maps \citep[$A_v$=8-70 mag,][]{schlegel98}. The distance range of 6.0 to 9.0 kpc aligns with previous study estimates.

The SED fitting tool was not applied to YSOs in the two MIR saturated regions due to the inadequacy of relying solely on NIR photometric data, which does not provide a sufficient basis for drawing reliable conclusions. This limitation impacted 217 YSOs. Consequently, we derived relatively robust parameters for 1,340 out of the 1,814 selected YSOs (64 Class\,I and 1,276 Class\,II) with a $\chi^2_{best}$<100, accounting for 74\% of the total sample. The inability to determine reliable parameters for the remaining 26\% of the objects can be attributed to several factors, including errors in the identification and selection process, as well as intrinsic challenges related to the data quality and completeness.

To assess the 217 objects located in MIR-saturated areas that lacked robust SEDs, we checked our final sample for artefacts using the k\_1ppErrBits parameter from the UKIDSS\,GPS survey \citep{lucas08, Solin2012}. Among them only for 2 k\_1ppErrBits = 64 (indicating bad pixel(s)) and for 5 k\_1ppErrBits = 65552 (signifying proximity to saturation). Additionally, we conducted a visual inspection of them. The analysis of the surface brightness distribution for 38 faint objects suggested that they were more likely to be extended/non-stellar rather than point sources. To ensure the purity of our sample, these 38 objects were removed from the final list.

Thus, in the final list, we included only 1,482 YSOs, comprising 1,313 with constructed SEDs and 169 YSOs in the two saturated regions. Coordinates, NIR and MIR photometric data, as well as the evolutionary stage of the stellar objects associated with IRAS sources are presented in Table \ref{tab:NIRIRAS}. Table \ref{tab:FIRIRAS} presents their FIR fluxes. Additionally, Table \ref{tab:SEDIRAS} displays the weighted means and standard deviations of their parameters for all models with $\chi^2-\chi^2_{best}<3N$ obtained by the SED fitting tool. Figure \ref{fig:SED} displays the constructed SEDs for objects associated with the IRAS sources. Comprehensive tables, containing the same parameters for all selected YSOs, \textbf{will be available in the \textit{VizieR} database}.

From the data obtained using the SED fitting tool, we can infer several conclusions. The average interstellar extinction is calculated to be $A_v\,\approx$\,11\,mag, which is on the lower end of the estimates obtained from COBE/DIRBE and IRAS/ISSA maps \citep{schlegel98}. The average mass of the YSOs is $\sim$\,4.0\,M$_{\odot}$, with the minimum estimated at 1.5\,M$_{\odot}$ and the maximum at 22\,M$_{\odot}$. The considerable distance of the star-forming region may account for the scarcity of identified low-mass stellar objects.

\begin{table*}
\caption[]{NIR and MIR photometric data of IRAS sources}
\resizebox{1\textwidth}{!}{
\label{tab:NIRIRAS}
\begin{tabular}{l l l *{12}{c} l}
\hline\hline
\noalign{\smallskip}
\centering
IRAS	&	 $\alpha$(2000)			&	$\delta$(2000)			&	J			&	H			&	K			&	[3.6]\,$\mu$ m			&	[4.5]\,$\mu$ m			&	[5.8]\,$\mu$ m			&	[8.0]\,$\mu$ m			&	[24]\,$\mu$ m			&	W1			&	W2			&	W3			&	W4			&	Class	\\
	&	(hh mm ss)  			&	(dd mm ss) 			&	(mag)			&	(mag)			&	(mag)			&	(mag)			&	(mag)			&	(mag)			&	(mag)			&	(mag)			&	(mag)			&	(mag)			&	(mag)			&	(mag)			&		\\
\hline\noalign{\smallskip}																							
(1) & (2)& (3)&(4) &(5)&(6)             &               (7)             &               (8)             &               (9)             &               (10)            &               (11)            &               (12)            &       (13) & (14) & (15) & (16)   \\
\hline \noalign{\smallskip}
19117+1107	&	19	14	9.4	&	11	13	0.5	&	$-$			&	$-$			&	16.07	$\pm$	0.05	&	13.81	$\pm$	0.17	&	12.85	$\pm$	0.17	&	$-$			&	$-$			&	$-$			&	$-$			&	$-$			&	$-$			&	$-$			&	I	\\
19120+1103	&	19	14	21.9	&	11	9	15.0	&	10.83	$\pm$	0.00	&	11.27	$\pm$	0.00	&	10.44	$\pm$	0.00	&	$-$			&	$-$			&	$-$			&	$-$			&	$-$			&	$-$			&	$-$			&	-2.24	$\pm$	0.00	&	-4.84	$\pm$	0.00	&	II	\\
19123+1101	&	19	14	40.6	&	11	6	56.5	&	19.30	$\pm$	0.15	&	16.10	$\pm$	0.02	&	14.45	$\pm$	0.01	&	13.49	$\pm$	0.13	&	12.93	$\pm$	0.14	&	$-$			&	$-$			&	$-$			&	$-$			&	$-$			&	$-$			&	$-$			&	II	\\
19123+1118	&	19	14	43.9	&	11	23	47.2	&	16.48	$\pm$	0.01	&	15.23	$\pm$	0.01	&	14.52	$\pm$	0.01	&	13.82	$\pm$	0.09	&	13.42	$\pm$	0.12	&	$-$			&	$-$			&	$-$			&	$-$			&	$-$			&	$-$			&	$-$			&	II	\\
19124+1106	&	19	14	47.3	&	11	11	34.8	&	18.69	$\pm$	0.08	&	15.12	$\pm$	0.01	&	13.34	$\pm$	0.00	&	12.06	$\pm$	0.07	&	11.76	$\pm$	0.11	&	$-$			&	$-$			&	$-$			&	$-$			&	$-$			&	$-$			&	$-$			&	II	\\
19124+1115	&	19	14	48.2	&	11	20	38.0	&	17.42	$\pm$	0.02	&	15.72	$\pm$	0.01	&	14.72	$\pm$	0.01	&	13.94	$\pm$	0.11	&	13.76	$\pm$	0.15	&	$-$			&	$-$			&	$-$			&	$-$			&	$-$			&	$-$			&	$-$			&	II	\\
19125+1057	&	19	14	51.8	&	11	2	54.0	&	16.55	$\pm$	0.01	&	15.05	$\pm$	0.01	&	13.78	$\pm$	0.01	&	12.61	$\pm$	0.30	&	12.07	$\pm$	0.19	&	$-$			&	$-$			&	$-$			&	$-$			&	$-$			&	$-$			&	$-$			&	II	\\
19125+1057.1	&	19	14	53.7	&	11	2	32.1	&	19.45	$\pm$	0.16	&	16.30	$\pm$	0.02	&	14.76	$\pm$	0.01	&	13.49	$\pm$	0.11	&	13.16	$\pm$	0.23	&	$-$			&	$-$			&	$-$			&	$-$			&	$-$			&	$-$			&	$-$			&	II	\\
19126+1110	&	19	14	57.1	&	11	16	17.5	&	17.21	$\pm$	0.02	&	15.19	$\pm$	0.01	&	14.22	$\pm$	0.01	&	13.13	$\pm$	0.09	&	12.65	$\pm$	0.16	&	$-$			&	$-$			&	$-$			&	$-$			&	$-$			&	$-$			&	$-$			&	II	\\
19126+1113	&	19	15	0.9	&	11	19	12.3	&	16.33	$\pm$	0.01	&	14.65	$\pm$	0.01	&	13.66	$\pm$	0.01	&	12.89	$\pm$	0.07	&	12.69	$\pm$	0.07	&	$-$			&	$-$			&	$-$			&	$-$			&	$-$			&	$-$			&	$-$			&	II	\\

\hline\noalign{\smallskip}
\end{tabular}
}

\textit{\textbf{Notes.}} (1) - name of the IRAS sources, (2),(3) - UKIDSS\,GPS coordinates of YSOs associated with IRAS sources, (4)–(15) - apparent magnitudes with errors (UKIDSS\,GPS (4)-(5), \textit{Spitzer} (7)-(11), WISE (12)-(15)), (16) - final classification of evolutionary stages according to c-c diagrams. 
\end{table*}

\begin{table*}
\caption[]{FIR photometric data of IRAS sources }
\resizebox{1\textwidth}{!}{
\label{tab:FIRIRAS}
\begin{tabular}{*{10}{l} }
\hline\hline
\noalign{\smallskip}
\centering
IRAS 	& [12]\,$\mu$m	&	[25]\,$\mu$m			&	[60]\,$\mu$m	&	[100]\,$\mu$m &	[70]\,$\mu$m			&	[160]\,$\mu$m			&	[250]\,$\mu$m			&	[350]\,$\mu$m			&	[500]\,$\mu$m		\\
		&	(Jy) 		&	(Jy) 		&(Jy) 			&(Jy) 		&(Jy) 			&(Jy) 			&(Jy) 			&(Jy) 			&(Jy) 		\\
\hline\noalign{\smallskip}
(1)     &               (2)             &               (3)                 &            (4)             &               (5)             &               (6)             &               (7)             &               (8)             &               (9) &(10)  \\

\hline \noalign{\smallskip}
19117+1107	&	37.3	$\pm$	2.24	&	304	$\pm$	15.2	&	2610	$\pm$	182.7	&	7890	$\pm$	3945	&	252	$\pm$	0.9		&	155		$\pm$	1.3	&	$-$			&	$-$			&	$-$			\\
19120+1103	&	78.8	$\pm$	3.94	&	640	$\pm$	32	&	5340	$\pm$	373.8	&	7890	$\pm$	867.9	&	1552	$\pm$ 1.6		&	842		$\pm$ 1.9		&	465 $\pm$ 1.6			&	$-$			&	96 $\pm$ 1.7		\\
19123+1101	&	1.34	$\pm$	0.17	&	0.83	$\pm$	0.15	&	31	$\pm$	15.50	&	7890	$\pm$	3945	&	$-$			&	2.1 $\pm$	0.6 	&	$-$			&	$-$			&	$-$			\\
19123+1118	&	1.67	$\pm$	0.84	&	1.80	$\pm$	0.18	&	13.8	$\pm$	2.35	&	63.5	$\pm$	31.75	&	$-$			&	$-$			&	$-$			&	$-$			&	$-$			\\
19124+1106	&	3.92	$\pm$	0.39	&	19.6	$\pm$	2.16	&	261	$\pm$	41.76	&	582	$\pm$	87.3	&	170		$\pm$	0.5 &	104	$\pm$	0.7	&	48 $\pm$	0.7		&	20 $\pm$	0.8			&	20 $\pm$	1.6		\\
19124+1115	&	1.21	$\pm$	0.10	&	1.06	$\pm$	0.12	&	7.28	$\pm$	3.64	&	582	$\pm$	291	&	$-$			&	1.6	$\pm$	0.2		&	1.5 $\pm$	0.5			&	$-$			&	$-$				\\
19125+1057	&	2.53	$\pm$	1.27	&	3.95	$\pm$	0.43	&	60	$\pm$	7.80	&	186	$\pm$	93	&	84	$\pm$	0.3	&	63 $\pm$	0.8		&	22 $\pm$	0.7		&	17 $\pm$	1.5			&	7.6 $\pm$	1.5		\\
19125+1057.1	&	2.53	$\pm$	1.27	&	3.95	$\pm$	0.43	&	60	$\pm$	7.80	&	186	$\pm$	93	&	84	$\pm$	0.3	&	63 $\pm$	0.8		&	22 $\pm$	0.7		&	17 $\pm$	1.5			&	7.6 $\pm$	1.5		\\
19126+1110	&	0.77	$\pm$	0.06	&	2.19	$\pm$	0.22	&	63.5	$\pm$	6.99	&	582	$\pm$	291	&	6.1		$\pm$	0.1 &	2.36	$\pm$	0.32	&	$-$			&	$-$			&	$-$			\\
19126+1113	&	1.72	$\pm$	0.86	&	0.76	$\pm$	0.11	&	7.61	$\pm$	3.81	&	582	$\pm$	291	&	3.3		$\pm$	0.1 &	2.0 	$\pm$	0.4 &	1.5 $\pm$	0.4		&	$-$			&	1.2 $\pm$	0.6			\\

\hline\noalign{\smallskip}
\end{tabular}
}

\textit{\textbf{Notes.}} (1) - name of the IRAS sources,  (2)–(10) - measured fluxes with errors (IRAS (2)-(5), \textit{Herschel} (6)-(10)). 
\end{table*}

\begin{table*}
\caption[]{Parameters of IRAS sources derived from \citet{robitaille07} models SED fitting.}
\resizebox{1\textwidth}{!}{
\label{tab:SEDIRAS}
\begin{tabular}{l l l *{4}{c} l *{4}{c} }
\hline\hline
\noalign{\smallskip}
\centering
IRAS	&	N$_{d}$ 	&	N$_{fit}$	&	A$_v$			&	Distance			&	Stellar age			&	Stellar mass			&	Temperature			&	M$_{disk}$			&	MdotE			&	MdotD			&	L$_{Total}$			\\
	&		&		&	(mag)			&	(kpc)			&	(Log)(yr)			&	(M$_{\odot}$)		&	(K)			&	 (Log)(M$_{\odot} $)			&	 (Log)(M$_{\odot} yr^{-1}$)			& (Log)(M$_{\odot} yr^{-1}$)			&	(Log)(L$_{\odot}$)					\\
\hline \noalign{\smallskip}														
\hline\noalign{\smallskip}
(1)     &               (2)             &               (3)                 &            (4)             &               (5)             &               (6)             &               (7)             &               (8)             &               (9)             &               (10)            &               (11)            &               (12)              \\

\hline \noalign{\smallskip}
19117+1107	& 6	&	2 &	49.7	$\pm$	1.6 &	6.2	$\pm$	1.1	&	4.9	$\pm$	4.7	&	13.7	$\pm$	0.4	&	28253 $\pm$	3232	&	-1.5	$\pm$	-3.4	&	-3.0	$\pm$	-3.8	&	-5.1	$\pm$	-5.4	&	4.3	$\pm$	3.5	\\
19120+1103	&	10	& 1 &	25.8	$\pm$	0.0	&	6.0	$\pm$	0.0	&	4.2	$\pm$	0.0	&	21.8	$\pm$	0.0	&	23880	$\pm$	0	&	$-$			&	-2.1	$\pm$	0.0	&	$-$			&	4.9	$\pm$	0.0	\\

19123+1101	&	8	&	3		& 17.4	$\pm$	0.6	&	7.3	$\pm$	1.2	&	5.5	$\pm$	5.0	&	7.2	$\pm$	0.6	&	17376	$\pm$	2563	&	-1.9	$\pm$	-2.1	&	-5.0	$\pm$	-5.7	&	-6.2	$\pm$	-6.1	&	3.4	$\pm$	3.0
\\
19123+1118	&	7	&	2	&	5.2	$\pm$	0.6	&	7.2	$\pm$	1.2	&	5.8	$\pm$	5.2	&	7.1	$\pm$	0.4	&	20769	$\pm$	412	&	-0.9	$\pm$	-1.0	&	-5.9	$\pm$	-5.8	&	-6.7	$\pm$	-7.1	&	3.3	$\pm$	2.9
	\\
19124+1106	&	12	& 1	& 6.9	$\pm$	0.0	&	8.8	$\pm$	0.0	&	4.3	$\pm$	0.0	&	15.5	$\pm$	0.0	&	13830	$\pm$	0	&	-2.8	$\pm$	0.0	&	-2.6	$\pm$	0.0	&	-5.9	$\pm$	0.0	&	4.3	$\pm$	0.0	\\

19124+1115	&	7	&	2	&	9.6	$\pm$	0.4	&	8.6	$\pm$	1.0	&	5.6	$\pm$	5.1	&	6.9	$\pm$	0.9	&	17238	$\pm$	6336	&	-2.2	$\pm$	-2.7	&	-5.4	$\pm$	-5.8	&	-7.0	$\pm$	-6.9	&	3.2	$\pm$	2.7
\\
19125+1057	&	12	& 1	& 5.8 $\pm$	0.0	&	7.8	$\pm$	0.0	&	3.9	$\pm$	0.0	&	13.7	$\pm$	0.0	&	5066	$\pm$	0	&	-2.0	$\pm$	0.0	&	-2.7	$\pm$	0.0	&	-7.3	$\pm$	0.0	&	3.9	$\pm$	0.0	\\

19125+1057.1	&	12	&	2	&	14.3	$\pm$	11.5	&	7.5	$\pm$	1.2	&	3.7	$\pm$	3.6	&	16.5	$\pm$	4.4	&	4441	$\pm$	366	&	-1.9	$\pm$	-1.7	&	-2.7	$\pm$	-3.1	&	$-$			&	3.9	$\pm$	3.5
	\\
19126+1110	&	10	& 1	& 10.0	$\pm$	0.0	&	6.0	$\pm$	0.0	&	5.7	$\pm$	0.0	&	7.4	$\pm$	0.0	&	21490	$\pm$	0	&	-2.3	$\pm$	0.0	&	-5.5	$\pm$	0.0	&	-8.8	$\pm$	0.0	&	3.3	$\pm$	0.0	\\

19126+1113	&	10	& 1	& 8.9	$\pm$	0.0	&	7.1	$\pm$	0.0	&	5.0	$\pm$	0.0	&	7.2	$\pm$	0.0	&	5908	$\pm$	0	&	-1.8	$\pm$	0.0	&	-4.6	$\pm$	0.0	&	-6.2	$\pm$	0.0	&	2.9	$\pm$	0.0	\\

\hline\noalign{\smallskip}

\end{tabular}
}
\textit{\textbf{Notes.}} (1) - name of the IRAS sources, (2) - number of input fluxes, (3) - number of fitting models, (4)-(12) - the weighted means and the standard deviations of parameters obtained by SED fitting tool for all models with best $\chi^{2}-\chi^{2}_{best}$\,<\,3N. 
\end{table*}

\begin{figure*}[h!t]
\includegraphics[width=0.9\linewidth]{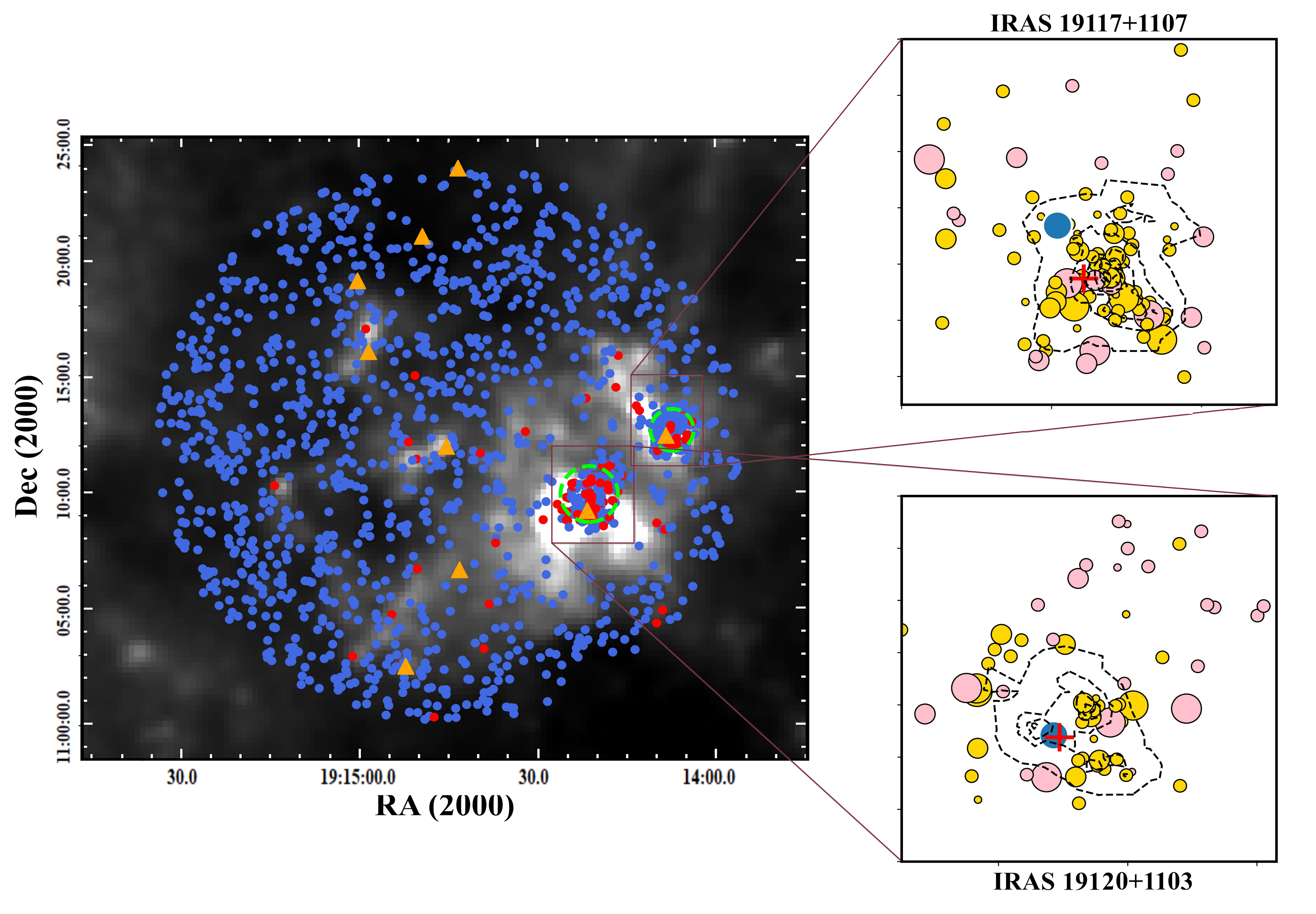}
    \caption{(\textit{Left panel}): Distribution of YSOs in the region on \textit{Herschel} 500\,$\mu$m image. Class\,I and Class\,II objects are indicated by filled red and blue circles, respectively. IRAS sources are marked with orange triangles. Dashed green circles denote the MIR-saturated regions. (\textit{Right panels}): The two insets show the distribution of the clusters’ members with fitted SEDs. Yellow and pink circles correspond to an older and younger population, respectively. The size of each circle is related to its mass falling within certain interval of masses: 1–3\,M$_\odot$ (smallest), 3–5\,M$_\odot$, 5–7\,M$_\odot$, >7\,M$_\odot$ (largest) obtained with SED fitting tool. Red crosses show the positions of IRAS\,19117+1107 and IRAS\,19120+1103, while blue circles represent  their NIR counterparts. Dashed contours represent the local density distribution of the YSOs within the clusters.}
    \label{fig:dis}
\end{figure*}


\subsection{Distribution of YSOs}
\label{sec:dis}

The left panel of Figure \ref{fig:dis} displays the distribution of the selected YSOs in the field. Class I and Class II objects are represented by filled red and blue circles, respectively. The IRAS sources are indicated by orange triangles and the dashed green circles shows the MIR-saturated regions. Apart from the regions surrounding the IRAS 19120+1103 and 19117+1107 sources, the YSOs of both evolutionary classes are quite uniformly distributed throughout the molecular cloud.  However, near both UC\,HII regions, the YSOs are notably clustered, forming dense concentrations, confirming the radial distribution findings presented in Figure \ref{fig:radial}. We determined the radius of each cluster from its geometric center based on the YSO density distribution. Table \ref{tab:IRAS} details the clusters' characteristics, such as geometric centers, radii, and stellar surface densities, which are significantly higher than the region's average. In the right panels of Figure \ref{fig:dis}, two insets highlight the distribution of cluster members with different masses and evolutionary stages for which SED analysis was conducted. Dashed contours illustrate the local density distribution of the members, both with and without SED analysis. \\

\textbf{IRAS\,19117+1107 (MSX\,G045.4782+00.1323).} Figure \ref{fig:dis} clearly depicts a cluster of YSOs predominantly concentrated around the IRAS source. Analysis of the stellar population revealed a massive YSO with 13.7\,$\pm$\,0.4\,M$_{\odot}$ mass and $\sim$\,30,000\,K temperature. The parameters of this stellar object are detailed in Tables \ref{tab:NIRIRAS} - \ref{tab:SEDIRAS} (\#116 in the online tables). This stellar object, however, is significantly distanced from the IRAS source ($\sim$\,30\,arcsec), and does not fall into the ellipse of uncertainty. Closer to the IRAS source, inside the uncertainty ellipse, there is another star, which, within the error bar, can be considered as a high-mass YSO (7.2\,$\pm$\,2.1\,M$_{\odot}$ mass and $\sim$\,20,000\,K temperature, \#57 in the online Tables). Additionally, five other YSOs with masses greater than 7\,M$_\odot$ have been identified in the cluster (\#28, 36, 91, 109, 115). It seems that the group of the above-mentioned high-mass stars constitute the ionization power source of G\,45.48+0.13. However, the substantial mass and temperature of object \#116 suggest that it plays a primary role in this process. \\

\textbf{IRAS\,19120+1103 (MSX\,G045.4543+00.0600).} Similar to the previous case, a cluster of YSOs has been identified here. Within this cluster, a massive YSO with 21.8\,M$_{\odot}$ mass and 24,000\,K temperature is located in the immediate vicinity of the IRAS source. The parameters of this YSO  are detailed in Tables\,\ref{tab:NIRIRAS} - \ref{tab:SEDIRAS} (\#261 in the online Tables). This massive YSO is associated with ammonia and water maser emissions \citep{Urquhart2011}. In the vicinity of this object, we identified two additional high-mass YSOs with masses of 11.2 and 9.7 M$_\odot$ (\#207 and 268, respectively). These YSOs, including the massive YSO \#261, are likely contributors to the formation of the HII region. Undoubtedly, as with the previously discussed case, the most massive object plays a dominant role. Our findings align with previous studies suggesting that the region is ionized by three O-type stars \citep{Paron2009} and that G45.45+0.06 is essentially a young OB cluster \citep{Feldt1998}. 

In the vicinity of G\,45.47+0.05, we identified three stars with a mass greater than 8\,M$_\odot$ (\#305, 307, and 311 in the online Table). One (or more) of these stars may be the candidates for driving the molecular outflows, as suggested in previous studies \citep{Ortega2012, Paron2013}.\\

\begin{table}
\caption[]{Properties of the region}
\resizebox{1\textwidth}{!}{
\label{tab:IRAS}
\begin{tabular}{l c cccc c c}
\hline\hline
\noalign{\smallskip}
\centering
Name & $\alpha$(2000) & $\delta$(2000) & $\alpha$(2000) & $\delta$(2000) & Radius & N & Density\\
& (hh mm ss) & (dd mm ss) & (hh mm ss) & (dd mm ss) & (arcmin) & & (arcmin$^{-2}$)\\
\hline\noalign{\smallskip}
(1) & (2) &  (3) & (4) & (5) & (6) & (7) & (8)\\
\hline \noalign{\smallskip}
IRAS\,19120+1103 & 19 14 20.8 & +11 09 04 & 19 14 20.6 & +11 09 55.2 & 1.9 & 172 & 15\\
IRAS\,19117+1107 & 19 14 08.6 & +11 12 27 & 19 14 07.3 & +11 12 39.67 & 1.4 & 143 & 23\\
Region & - & - & 19 14 45.0 & +11 12 06 & 12 & 1482 & 3.2\\
\hline\noalign{\smallskip}
\end{tabular}
}

\textit{\textbf{Notes.}} (1) - names of (sub-)regions, (2),(3) - coordinates of the IRAS sources, (4),(5) - coordinates of the geometric centres, (6) - radii of the (sub-)regions, (7) - numbers of objects within selected radii, (8) - surface stellar density in the (sub-)regions.
\end{table}

Several additional IRAS sources are located within the study area. Surrounding these sources, we did not identify regions with increased surface stellar density. However, stars with high and intermediate masses are associated with these sources (see Tables\,\ref{tab:NIRIRAS} - \ref{tab:SEDIRAS}, objects \#740, 780, 825, 839, 895, 933, 985, and 1044 in online Tables). Among these, IRAS 19124+1106, associated with MSX6C\, G045.5426-00.0067 source, is particularly noteworthy. It has been postulated to be associated with the termination of the GRS\,1915+105 jet \citep{Rodriguez1998}, though this hypothesis has been later questioned \citep{Zdziarski2005}. There is no information about other sources in previous studies. Four of the remaining IRAS sources (19123+1101, 19124+1115, 19126+1110, and 19126+1113) are associated with MSX6C sources G045.4518-00.0286, G045.6826+00.0493, G045.6231-00.0217, and G045.6756-00.0050, respectively.

\begin{figure*}[h!t]
\includegraphics[width=0.8\linewidth]{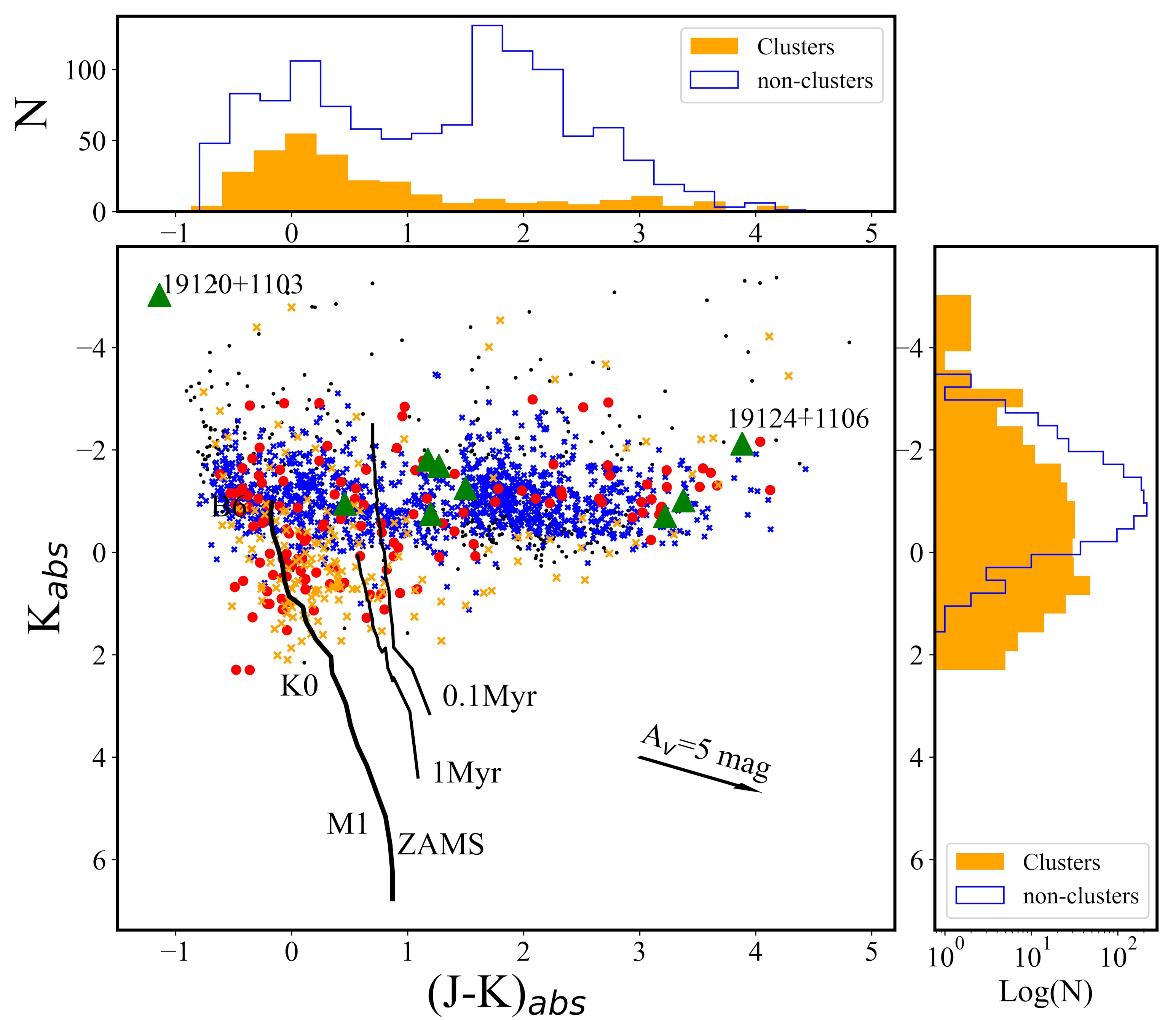}
\caption{K vs. (J–K) colour-magnitude diagram for identified YSOs in the considered region (\textit{bottom left panel}). The PMS isochrones for 0.1 and 1 Myr \citep{siess00} and the ZAMS are represented by solid thin and thick lines, respectively. Several spectral types are labeled. The J and K magnitudes of YSOs have been corrected for interstellar extinction using an average A$_v$\,=\,11\,mag value, determined by the SED fitting tool. Red circles denote stellar objects within the IRAS clusters with constructed SEDs. Objects in the saturated regions are marked with yellow crosses, while non-cluster objects are indicated by blue crosses. The IRAS sources are shown as green triangles, with two of them labeled. Black dots are objects without reliable parameters.(\textit{Top panel}): Histograms of (J-K)$_{abs}$ values.  (\textit{Right panel}): KLFs for the clusters's members and non-cluster objects.}
    \label{fig:CMD}
\end{figure*}

\begin{figure}
    \includegraphics[width=0.9\linewidth]{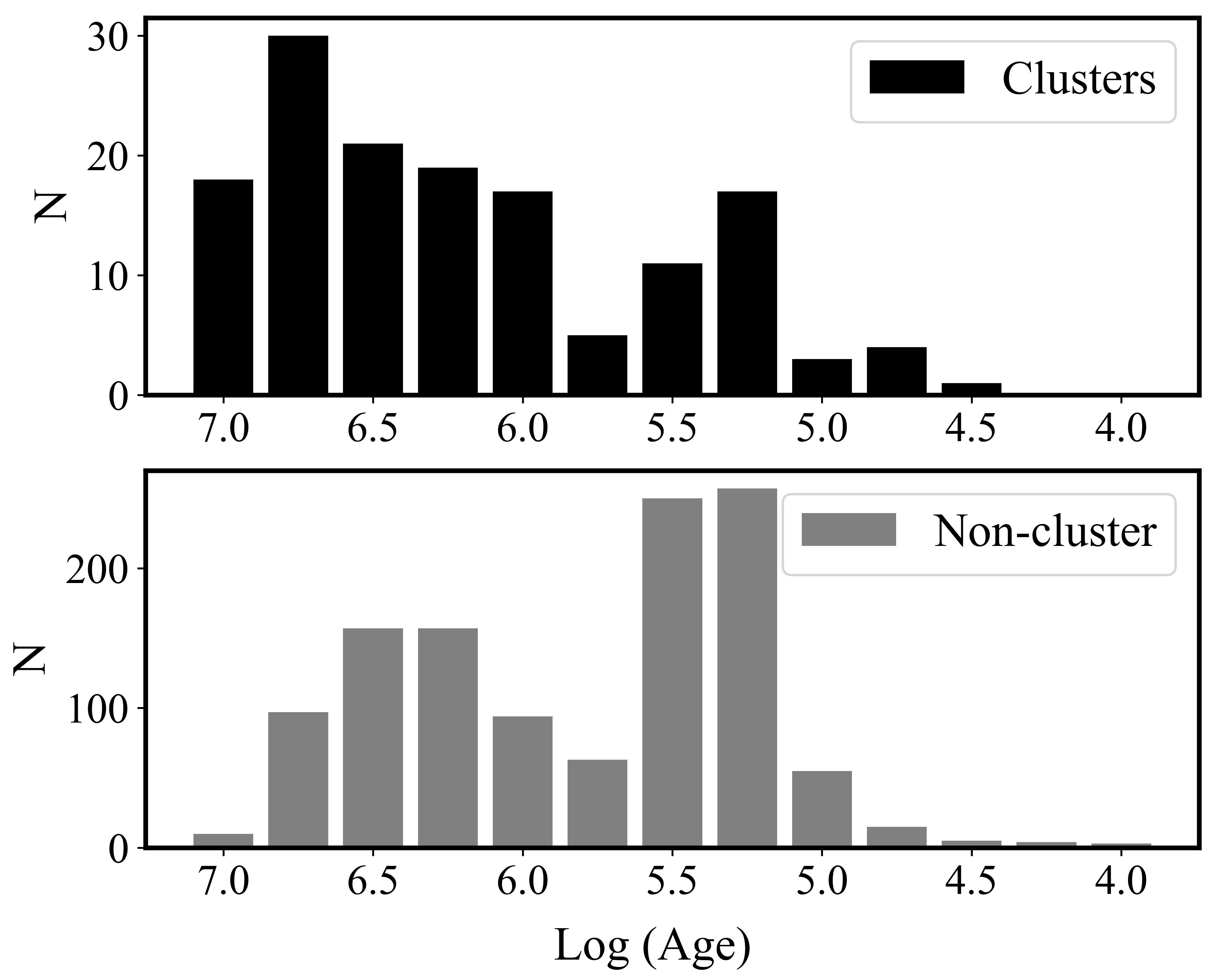}
    \caption{Histogram of evolutionary ages (by the SED fitting tool) for members of the IRAS clusters (\textit{top panel}) and the non-cluster objects (\textit{bottom panel}). The bin size corresponds to Log\,(Age)\,=\,0.25.}
    \label{fig:Age}
\end{figure}

\subsection{Colour-magnitude diagram and evolutionary age spread} 
\label{sec:cmd}

A powerful method for analyzing the characteristics of stellar populations in star-forming regions is the colour-magnitude diagram (CMD). As in Paper I, we used a K versus J–K CMD (Figure \ref{fig:CMD} lower left panel) to display distribution of the identified YSOs. The Zero Age Main Sequence (ZAMS) (thick solid curve) and the Pre-Main Sequence (PMS) isochrones for ages 0.1 and 1\,Myr  (thin solid curves) are adopted  from \citet{siess00}, using bolometric $M_J$ and $M_K$ magnitudes. We utilized the conversion table from \citet{kenyon94}. Stars within the IRAS clusters, for which the SED fitting tool yielded reliable parameters ($\chi^2$<100), are represented by red circles. Yellow crosses denote objects in the saturated regions, limited to three photometric measurements. Blue crosses mark the non-cluster objects found outside the IRAS clusters. Black dots correspond to objects with unreliable parameters ($\chi^2$>100). The J and K magnitudes of the YSOs were adjusted for a distance of 7.8\,kpc and the average interstellar extinction ($A_v$\,=\,11\,mag, see Section \ref{subsec:selection}). We did not fully consider that star-forming regions typically have a significant gradient of interstellar extinction. This variation may cause some displacement of an object relative to isochrones and $K_{abs}$. However, for any given object, the value of $A_v$ can vary, being either higher or lower than the average. Consequently, the overall distribution of the stellar population on the CMD still presents a realistic picture.

Black dots in the diagram are presumably foreground objects. The YSOs within the star-forming region (circles and crosses) show substantial variations in their (J-K)$_{abs}$ values. Meanwhile, the members of the IRAS clusters and non-cluster objects are differently distributed. The non-cluster objects are grouped into two main concentrations distributed on the right and left of the 0.1\,Myr isochrone. Conversely, $\sim$\,75\% of objects in the IRAS clusters are situated to the left of the 0.1\,Myr isochrone and are mainly clustered around the ZAMS. For enhanced clarity, we have included histograms of (J-K)$_{abs}$ values, which serve as a rough indicator of the evolutionary stage or age of YSOs (top panel of Figure \ref{fig:CMD}). Generally, both samples differ significantly in their distribution. Most IRAS cluster members appear bluer, suggesting a more advanced evolutionary stage, whereas the non-cluster objects are divided into two distinct groups, with the bluer group overlapping with the IRAS cluster members. However, both groups are evenly dispersed across the field.

The middle panel of Figure \ref{fig:Age} illustrates the distribution of evolutionary ages, as determined by the SED fitting tool, for the non-cluster objects. This confirms our previous findings. The age distribution of these objects features two distinct peaks at Log(Age[years])\,$\approx$\,6.25 and 5.25. The second peak coincides with the peak of the non-cluster objects in G\,45.07+0.13 and G\,45.12+0.13 star-forming regions, which are belonging to the same molecular cloud (see Fig.\,10 in Paper\,I). This consistency confirms that these objects are indeed part of the young stellar population within the GRSMC\,45.46+0.05 molecular cloud. 

In contrast, the evolutionary age distribution of the cluster members (Fig. \ref{fig:Age}, top panel) exhibits a primary peak at Log(Age)\,$\approx$\,6.75, and a minor peak at Log(Age)\,$\approx$\,5.25. This distribution is derived from parameters of only 146 YSOs, for which the SED fitting tool was applied. The majority of the remaining 169 YSOs, located in the MIR-saturated regions, are concentrated around the ZAMS and to the left of the 1\,Myr isochrone. This suggests that these objects significantly contribute to the first peak of the evolutionary age distribution. Therefore, the distribution will primarily feature a single well-defined peak at Log(Age)\,$\approx$\,6.75\,years, similar to what is observed in the neighboring G\,45.07+0.13 and G\,45.12+0.13 regions (see Paper\,I). Given the small scatter of the cluster members relative to the isochrones and their narrow age spread, it is plausible that the clusters originated from an external triggering shock.

We can also estimate the evolutionary age using fraction of NIR excess stars in the clusters, since the disc and/or envelope surrounding a PMS star becomes optically thinner with age \citep{Lada2003,Vig14}. For young clusters with an age $\sim$\,1\,Myr, the fraction of NIR excess stars is found to be $\sim$50$-$65\% \citep{Lada00,Muench01}. In clusters aged 1–2\,Myr, the NIR fraction is estimated to be $\sim$40\% \citep{Kenyon95, Vig14}, decreasing to about 20\% in older clusters aged 2–3\,Myr \citep{Teixeira05}. For our analysis, we estimated the fraction of field stars located close to the IRAS clusters, ensuring that the interstellar extinction effect on the fraction of background objects is almost the same. After correcting for the foreground and background star contamination, the fraction of the NIR excess stars in the clusters is estimated to be $\sim$50\%, indicating a lower age limit of 1\,Myr.

\subsection{K luminosity function}
\label{sec:klf}

Examining the slope $\alpha$ of the K-band luminosity function (KLF), where $dN(K)/dK \propto 10^{\alpha K}$, provides insights into the properties of the stellar population, including the Initial Mass Function (IMF), star formation history, and age distribution \citep{lada96}. The $\alpha$ slopes for the stellar populations within the IRAS clusters and for the non-cluster entities were independently assessed. This was achieved by applying a linear least-squares method to the counts of YSOs grouped in 0.5\,mag increments. The right panel of Figure \ref{fig:CMD} presents the superimposed KLFs for the IRAS clusters in orange and the non-cluster objects in blue. The lack of a pronounced drop at the KLF's tail suggests a smooth transition, indicative of the survey’s photometric completeness not substantially altering the shape of the KLF. Notably, the non-cluster objects exhibited a significantly sharper slope than the IRAS clusters, underscoring their different characteristics within the region and supporting the suggestion that they are at disparate evolutionary phases. The slopes for the KLFs were calculated to be 0.32$\pm$0.04 for the IRAS clusters and 0.72$\pm$0.13 for the non-cluster objects. The value of the power-law slope for the IRAS clusters is compatible to the typical values (0.32–0.38) for other young clusters \citep{Carpenter93,Lada93,Jose12}. Recent studies indicate that the KLF slope varies from 0.2 to 0.4 for clusters younger than 5\,Myr \citep{Blum00,Devine08}. This is compatible with the age estimate obtained earlier by the fraction of the NIR excess stars. The $\alpha$ slope value for the non-cluster objects is significantly steeper (see also Paper\,I), however, it is still in the range  of 0.18-0.94 value for other young clusters \citep{Carpenter93}.

\begin{figure}
\includegraphics[width=0.9\linewidth]{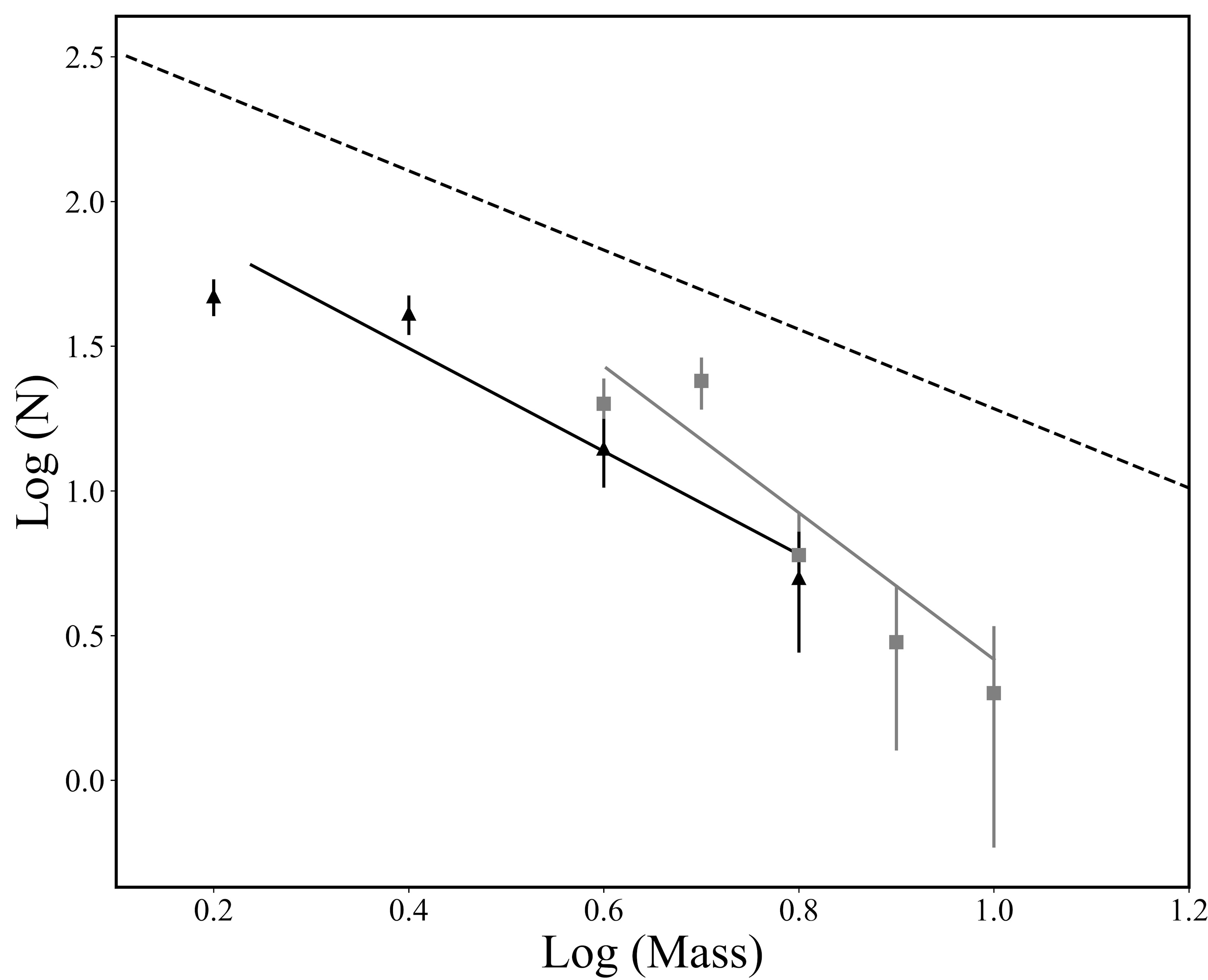}
\caption{Mass function of the IRAS clusters. Log (N) represents log(N/d log M). Grey squares and black triangles are the distributions of masses in 2.5$-$10\,M$_{\odot}$ and 1.5$-$10\,M$_{\odot}$ ranges, respectively. The error bars represent $\pm \sqrt{N}$ errors. The grey line shows a least-square fit to the mass range 2.5$-$10\,M$_{\odot}$, based on the SED fitting results. The black line shows a least-square fit to the mass range 1.5$-$10\,M$_{\odot}$ using the J vs. (J-H) CMD results. The dashed line represents mass function having Salpeter value $\Gamma$ = -1.35.}
\label{fig:IMF}
\end{figure}

\begin{figure}
\includegraphics[width=0.9\linewidth]{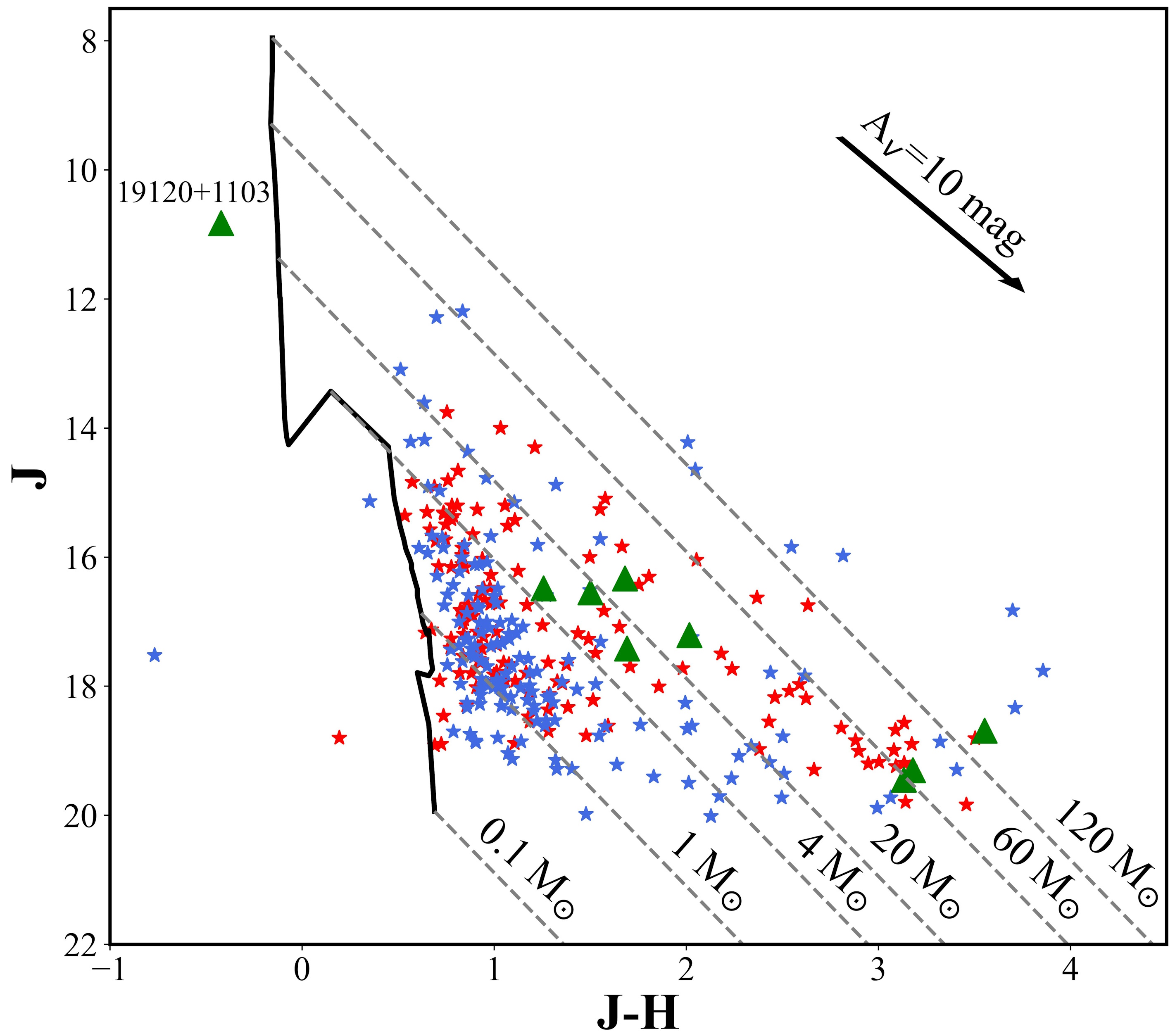}
\caption{J vs. (J-H) colour-magnitude diagram for the YSOs in the clusters. The black curve is the 1\,Myr PMS isochrone from \citet{siess00} and \citet{Lejeune01}. The reddening vectors (parallel dashed lines) for the 1\,Myr isochrone are drawn at 0.1, 1, 4, 20, 60 and 120\,M$_{\odot}$. The red stars represent the sources for which SED analysis was done, and the blue stars - without SED analysis. The IRAS sources are indicated by green triangles and one of them is labelled.}
\label{fig:CMDJJH}
\end{figure}

\begin{figure*}[h!t]
\includegraphics[width=0.9\linewidth]{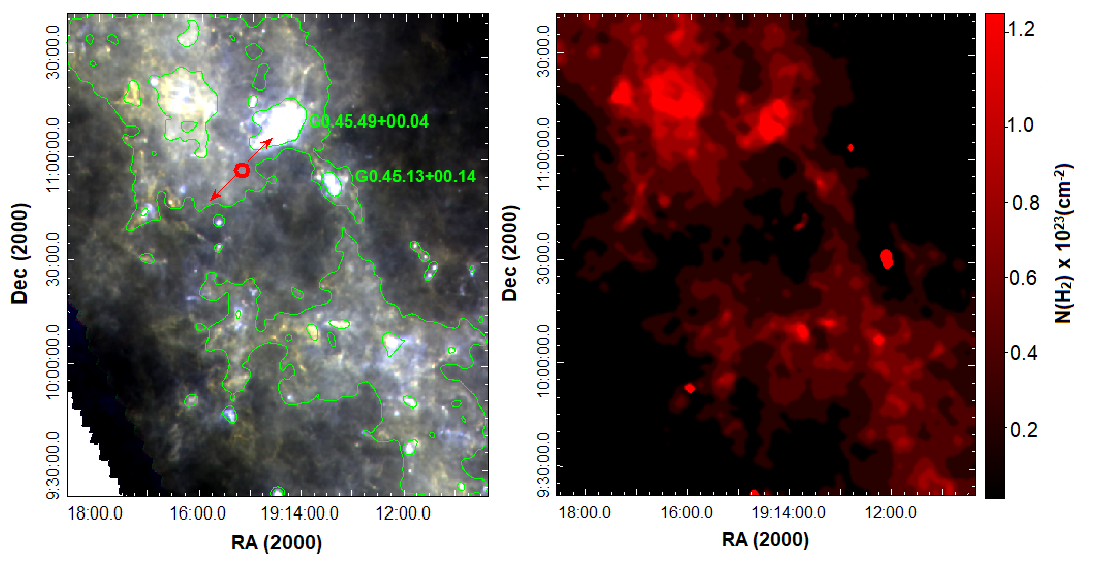}
\caption{\textit{(Left panel)}: colour-composite \textit{Herschel} image of GRSMC\,45.46+0.05 molecular cloud: 160\,$\mu$m (blue), 350\,$\mu$m (green), and 500\,$\mu$m (red) bands. The position of GRS\,1915+105 is marked with red circle and direction of its jet is marked with the red vectors. The green contours correspond to 0.15\,Jy/beam for 870\,$\mu$m emission. \textit{(Right panel)}: N(H$_2$) distribution map.  }
\label{fig:GRS}
\end{figure*}

\subsection{Mass function}
\label{sec:IMF}

Star-forming regions are crucial to study the IMF since their MFs can be essentially considered as IMFs. This is due to the fact that the regions are too young to have lost a significant number of members due to the effects of dynamical or stellar evolution \citep{mallick15}. Assuming that both the MF and the mass-luminosity relationship for a stellar cluster are described by power laws similar to the KLF, the slope of the KLF will be expressed as: $\alpha$=$\gamma/(2.5\beta)$, where $\gamma$ and $\beta$ are the slopes of MF and mass-luminosity relation, respectively. We discuss the MF only for the IRAS clusters. If we use $\beta$\textasciitilde2, generally used for a larger and higher mass range (O$-$F stars) at 1\,Myr \citep{balog04} then the calculated MF slope is $\gamma$\,=\,1.6$\pm$0.2, which is a slightly steeper than the Salpeter slope of 1.35 \citep{salpeter55} and close to the Miller-Scalo slope of 1.7 \citep{miller79, scalo86}.

We estimated the slope $\gamma$ using the SED fitting results. Given that most of our sources have masses within the range of 2.5$-$10\,M$_{\odot}$, we consider only this mass range for our analysis. The MF of the clusters is plotted in Figure \ref{fig:IMF} as a grey line. For the mass range 2.5$-$10\,M$_{\odot}$, the slope of the MF is determined to be $\gamma$\,=\,2.38$\pm$0.58. While this result aligns with the previously obtained $\gamma$ value only within its error bar, such a discrepancy could be attributed to the potential loss of members due to MIR saturation, which is leading to underestimation or non-detection of certain members, particularly those with high brightness in the MIR spectrum.

In addition to the results of the SED fitting tool, we also use the J vs. (J-H) CMD to estimate the mass range of the IRAS clusters. We avoid using a CMD involving the K-band magnitude, as this band is the most affected by the NIR excess flux from circumstellar material. Such excess flux can cause apparent brightening of sources, resulting an inaccurate mass estimates \citep{mallick15}. Figure \ref{fig:CMDJJH} presents the J vs. (J-H) CMD. On this diagram, the 1\,Myr PMS isochrone for masses between 0.1 to 7\,M$_{\odot}$, taken from \citet{siess00}, and for higher mass stars - from \citet{Lejeune01}, is depicted as a black curve. Reddening vectors for the 1\,Myr PMS isochrone are illustrated for masses of 0.1, 1, 4, 20, 60 and 120\,M$_{\odot}$. In the diagram the stars symbols represent members of the IRAS clusters; those analyzed using SED are marked in red, and those without SED analysis are in blue. The wide variation in colours of the YSOs is probably an indication of variable extinction, as well as different evolutionary stages of the sources. This may cause some sources to fall at the far-right end of the diagram. Major causes of uncertainty in this analysis include: (i) uncertainty in distance estimate, which is used to obtain the apparent magnitude (for the isochrone); (ii) unresolved binarity (especially since the source distance of $\sim$\,7.8\,kpc is quite large). Furthermore, the use of a specific PMS model introduces its own systematic error.

We estimated the slope $\gamma$ of the MF using the results from J vs. (J-H) CMD. As most of our sources fall within 1.5–10\,M$_{\odot}$ range, therefore, we considered only this mass range. The calculated slope of the MF within this range to be $\gamma$\,=\,1.69$\pm$0.34. The black curve in Figure \ref{fig:IMF} shows the fitted function for 1.5–10\,M$_{\odot}$ range. Notably, this value aligns more closely with the Salpeter slope (represented by the dashed line in the figure) and the Miller-Scalo slope than the slope $\gamma$ obtained from the SED fitting results.

We also compared our estimated values for $\alpha$ and $\gamma$ with those reported in other studies for a cluster aged 1\,Myr. \citet{balog04} obtained a value of $\gamma \sim$\,1.58 using $\beta$\,=\,2. \citet{Sanchawala07} obtained $\gamma$\,=\,1.30\,$-$\,1.40 value for 1\,Myr Trumpler clusters in the Carina Nebula. Given that the slopes of the KLF and MF of the IRAS clusters are similar to those of clusters aged  1\,Myr for a given mass-to-luminosity relation, it suggests that the IRAS clusters are likely around 1\,Myr old. This conclusion is consistent with the age estimate obtained earlier in this study.

\subsection{Is GRS\,1915+10 a trigger for star formation?}
\label{sec:Trigger}

In Sec. \ref{sec:cmd} we demonstrated that the distribution of the evolutionary ages of vast majority of the clusters' members  has a narrow spread (less than one order) with a well defined peak at Log(Age)\,$\approx$\,6.75 (see Fig. \ref{fig:Age}). Similarly, the members of the clusters in the neighboring G\,45.07+0.13 and G\,45.12+0.13 regions, which are physically connected with G\,045.49+00.04 \citep{Bhadari2022}, also exhibit the same distribution of ages (see Fig. 10 in Paper\,I). Furthermore, the majority of the clusters' members in all regions display a small scatter relative to the isochrones (see Fig. \ref{fig:CMD} here and Fig. 9 in Paper\,I). This pattern suggests that these clusters were likely formed almost simultaneously, implying that their origin may be attributed to an external triggering shock. A similar result was obtained and described in detail by \citet{Preibisch2008} for Upper Scorpius young stellar cluster, where the star formation process was hypothesized to have been triggered by the shock wave from an expanding superbubble, resulting in a young stellar population with an evolutionary age of around 5\,Myr and a narrow age spread.

There is substantial evidence suggesting that during the supersonic expansion of an ionized ISM, a dense layer of material can be collected between the ionization and the shock fronts \citep[see e.g.,][]{Elmegreen1977}. This layer might then fragment into massive condensations, potentially triggering the massive star formation \citep[see e.g.,][and ref. therein]{Pomares2009}. This star formation mechanism is known as the `collect and collapse' process. Indications that we are observing this process in action include the presence of a dense molecular shell surrounding the ionized gas in an HII region or massive fragments regularly positioned along the ionization front \citep[see e.g.,][]{Deharveng2009}. Consequently, HII regions often appear as rings in infrared imagery or as `bubbles' in projection, creating distinctive formations that have become a central focus in numerous studies investigating the triggering of star formation.

Figure \ref{fig:GRS} shows the colour-composite \textit{Herschel} image of GRSMC\,45.46+0.05 molecular cloud (160, 350, and 500\,$\mu$m), where the morphology of the ISM distinctly reveals a bubble shape. The G\,045.49+00.04 and G\,045.14+00.14 regions, located on the edge of this bubble, are clearly distinguished by their brightness. This bubble-like shape is particularly highlighted by the distribution of the hydrogen column density ($N(H_2$)). To derive $N(H_2)$, we used four intensity \textit{Herschel} maps ranging from 160 to 500 $\mu$m. The dust emission in the FIR range can be modelled as a modified blackbody, expressed as $I_{\nu}=k_{\nu}\mu_{H_{2}}m_{H}N(H_{2})B_{\nu}(T_d)$, where $k_{\nu}$ is the opacity, $\mu=2.8$ is the mean molecular weight, $m_{H}$ is the mass of hydrogen, and $B_{\nu}(T_d)$ is the Planck function at the dust temperature $T_d$. This procedure was detailed in Paper I. In the inner region of the bubble, the matter density is the lowest, with $N(H_{2})$ $\approx$ 10$^{21}$\,cm$^{-2}$. Moving toward the edge, $N(H_{2})$ increases up to $\sim$\,10$^{23}$\,cm$^{-2}$. The highest value of $N(H_{2})$, $\sim$\,3$\times$10$^{23}$\,cm$^{-2}$, is observed in the G\,045.49+00.04 region.

As mentioned in Sec. \ref{sec:introduction} within the GRSMC\,45.46+0.05 molecular cloud, and almost at the same distance (8.6$_{-1.6}^{+2.0}$ kpc), the GRS\,1915+105 X-ray binary is located, which contains a black hole and a K-giant companion (see Fig. \ref{fig:GRS}). The near-zero parallax of this system in relation to G045.07+0.13 indicates that this binary system did not receive a substantial velocity “kick”, due to mass loss from the primary as it evolved to a black hole \citep{Wu2019}. Black-hole X-ray binaries in a quiescent state are typically characterized by a luminosity on the order of L\,$\sim$\,10$^{34}$\,erg\,s$^{-1}$ with relatively short outbursts, during which their luminosity can reach $\sim$\,10$^{39}$\,erg\,s$^{-1}$. During these outbursts, these systems show clear repeating patterns of behavior associated with mechanical feedback in the form of winds and relativistic jets \citep{Ponti2012}. \citet{Fender2000} suggested that GRS\,1915+105 might inject more energy and matter into the outflow during periods of repeated small events compared to larger ejections. Consequently, the activity of this source, expected to last for at least several 10$^5$ years, will likely inject energy into the surrounding medium totaling at least approximately $\sim$\,10$^{51}$\,ergs. This amount of energy output is comparable to that of an average supernova event, significantly impacting the surrounding ISM.

Based on the above, several assumptions can be made about the process of star formation in G\,045.49+00.04 and G\,045.14+00.14 regions. Firstly, the ionized fronts in these regions are likely a result of the activity of GRS\,1915+105. The formation of young clusters near the IRAS sources likely was triggered by the shock front within the massive ISM condensation, through a "collect and collapse" process. In contrast, most non-cluster stellar objects in both regions are concentrated around the 0.1\,Myr isochrone, and their evolutionary age distribution has a well defined peak  at Log(Age)\,$\approx$\,5.25. These objects may represent a subsequent generation of stars, potentianlly formed due to the recurrent activity of GRS\,1915+105. As a complementary or alternative mechanism for the formation of these younger YSOs, one can also consider an `edge collapse' scenario, a process characterized by mass accumulation in each hub, facilitated by gas flows along filaments, as suggested in \citet{Bhadari2022}.

\section{Conclusion}
\label{sec:con}

The results of identification and study of the young stellar population in the molecular cloud G\,045.49+00.04, which includes three UC\,HII regions: G\,45.48+0.13 (IRAS 19117+110), G\,45.45+0.06 (IRAS 19120+1103), and G\,45.47+0.05, are presented. We used NIR, MIR and FIR photometric data, as well as the SED fitting tool to study YSOs. To determine the distribution of hydrogen column density, we applied the modified blackbody fitting on the \textit{Herschel} images obtained in four bands: 160, 250, 350, and 500\,$\mu$m. In this research the following results were obtained:

\begin{itemize}
    \item In total, we have identified 1,482 YSOs in a 12 arcmin radius covering GRSMC\,045.49+00.04, which include 1,382 Class\,II and 100 Class\,I objects. The average value of interstellar extinction is equal to A$_v\,\approx 11$\,mag. The maximum of YSOs' mass is estimated at 21.8 M$_{\odot}$ and the minimum is at 1.5 M$_{\odot}$. Most likely, large distance of the star-forming region can be responsible for the lack of low-mass stellar objects.
    \item The YSOs form relatively dense clusters in both UC\,HII regions, located close to the IRAS\,19120+1103 and 19117+1107 sources. The stellar density in the clusters is significantly higher than that of the entire region.  
    \item In all UC\,HII regions, several high-mass stars have been identified, which, in all likelihood, are sources of the ionization energy. The YSOs with 21.8\,M$_{\odot}$ and 13.7\,$\pm$\,0.4 M$_{\odot}$ are associated with IRAS\,19120+1103 and 19117+1107, respectively.  
    \item The non-cluster YSOs (1,168) are uniformly distributed in the field.
    \item The distribution of stellar objects from both samples on the CMD differs significantly. About 75\% objects of the IRAS clusters are concentrated around the ZAMS. On the contrary, the non-cluster objects have two concentrations located to the right and left of the 0.1 Myr isochrone. The bluer concentration coincides with members of the IRAS clusters.
    \item The distribution of evolutionary age of YSOs correlates well with their position on the CMD. The non-cluster YSOs have two well-defined peaks at Log(Age)\,$\approx$\,6.25 and 5.25. The evolutionary age distribution of the clusters’ members have well-defined peak at age Log(Age)\,$\approx$\,6.75, with a narrow spread (less than one order). In our previous study, it was shown that the YSOs in the neighboring G\,45.14+00.14 region have almost the same  distribution of evolutionary age. 
    \item The fraction of the NIR excess stars in the clusters is estimated to be $\sim$\,50\%, suggesting a lower age limit of 1\,Myr. A similar conclusion regarding age was drawn from the MF analysis. The calculation of the KLFs' $\alpha$ slopes for the IRAS clusters and non-cluster objects yielded 0.32$\pm$0.04 and 0.72$\pm$0.13, respectively. Thus, as observed in G\,45.14+00.14, the $\alpha$ slope for the non-cluster objects is significantly steeper, indicating that they are less evolved. This is well consistent with the above. 
    \item The colour-composite \textit{Herschel} image and the distribution of the $N(H_{2})$ reveal that the ISM morphology forms a bubble shape around GRS\,1915+105 (a black hole and a K-giant companion), and which, with a high probability, belongs to the same molecular cloud. G\,045.49+00.04 and G\,045.14+00.14 regions are located at the edge of this bubble, and are clearly distinguished by their brightness and density.
\end{itemize}

Based on the above, we can assume that the process of star formation in the young IRAS clusters was triggered by the GRS\,1915+105-initiated shock front inside the ISM massive condensation, through the "collect and collapse" process. Most non-cluster objects probably belong to a later generation. Their formation could be triggered by the recurrent activity of GRS\,1915+105 and/or through the edge collapse scenario and mass accumulation through the gas flows along the ISM filaments. Undoubtedly, this assumption about the star formation process requires further confirmation, including the study of the stellar population and the kinematic properties of the ISM in the GRSMC\,45.46+0.05 molecular cloud as a whole.

\section*{Acknowledgements} We thank the anonymous reviewer for constructive comments and suggestions. This work partially was made possible by a research grant number №\,21AG-1C044 from Science Committee of Ministry of Education, Science, Culture and Sports RA. This research has made use of the data obtained at UKIRT, which is supported by NASA and operated under an agreement among the University of Hawaii, the University of Arizona, and Lockheed Martin Advanced Technology Center; operations are enabled through the cooperation of the East Asian Observatory. We gratefully acknowledge the use of data from the NASA/IPAC Infrared Science Archive, which is operated by the Jet Propulsion Laboratory, California Institute of Technology, under contract with the National Aeronautics and Space Administration. We thank our colleagues in the GLIMPSE and MIPSGAL Spitzer Legacy Surveys. This publication also made use of data products from Herschel ESA space observatory.



\printbibliography

\end{document}